\documentclass[%
aip,%
floatfix,
reprint,%
onecolumn,
twocolumn,%
jcp,%
longbibliography,
]{revtex4-2}

\usepackage{graphicx}
\usepackage{dcolumn}
\usepackage{bm} 
\usepackage{color}

\usepackage[english]{babel} 

\usepackage{amsmath,amsthm,latexsym,amssymb,amsfonts,epsfig}
\usepackage{mathtools}
\usepackage{mathrsfs} 

\usepackage[outer=1.6cm, inner=1.6cm, top=2.2cm, bottom=2.2cm]{geometry}  

\usepackage{multirow}

\usepackage{ctable}

\usepackage{xcolor}
\definecolor{OliveGreen}{rgb}{0,0.6,0}

\definecolor{lightgray}{rgb}{0.83, 0.83, 0.83}

\newcommand{\vect}[1]{\boldsymbol{#1}}

\newcommand{\R}{\vect{r}}

\def\Xint#1{\mathchoice
   {\XXint\displaystyle\textstyle{#1}}%
   {\XXint\textstyle\scriptstyle{#1}}%
   {\XXint\scriptstyle\scriptscriptstyle{#1}}%
   {\XXint\scriptscriptstyle\scriptscriptstyle{#1}}%
   \!\int}
\def\XXint#1#2#3{{\setbox0=\hbox{$#1{#2#3}{\int}$}
     \vcenter{\hbox{$#2#3$}}\kern-.5\wd0}}

\def\dashint{\Xint-}

\usepackage[colorlinks=true,citecolor=OliveGreen,linkcolor=OliveGreen]{hyperref}

\usepackage{epstopdf}

\usepackage{csquotes}

\usepackage{type1ec} %

\usepackage{colortbl}

\DeclareMathAlphabet\mathbfcal{OMS}{cmsy}{b}{n}

\allowdisplaybreaks

\emergencystretch=\maxdimen
\begin{document}
	
	
	
	\title{Hydrodynamics of a disk in a thin film of weakly nematic fluid subject to linear friction}
	
	\author{Abdallah Daddi-Moussa-Ider}
	\email{abdallah.daddi-moussa-ider@open.ac.uk}
	\thanks{Author to whom any correspondence should be addressed.}

	\affiliation{School of Mathematics and Statistics, The Open University, Walton Hall, Milton Keynes MK7 6AA, UK}
	
	\author{Elsen Tjhung}
	
	\affiliation{School of Mathematics and Statistics, The Open University, Walton Hall, Milton Keynes MK7 6AA, UK}

	\author{Thomas Richter}
	
	\affiliation{Institut f\"ur Analysis und Numerik, Otto-von-Guericke-Universit\"at Magdeburg, Universit\"atsplatz 2, 39106 Magdeburg,
		Germany}

	\author{Andreas M. Menzel}
	\email{a.menzel@ovgu.de}
	
	\affiliation{Institut f\"ur Physik, Otto-von-Guericke-Universit\"at Magdeburg, Universit\"atsplatz 2, 39106 Magdeburg, Germany}

	\date{\today}

	\begin{abstract}
		To make progress towards the development of a theory on the motion of inclusions in thin structured films and membranes, we here consider as an initial step a circular disk in a two-dimensional, uniaxially anisotropic fluid layer. We assume overdamped dynamics, incompressibility of the fluid, and global alignment of the axis of anisotropy. Motion within this layer is affected by additional linear friction with the environment, for instance, a supporting substrate. We investigate the induced flows in the fluid when the disk is translated parallel or perpendicular to the direction of anisotropy. Moreover, expressions for corresponding mobilities and resistance coefficients of the disk are derived. Our results are obtained within the framework of a perturbative expansion in the parameters that quantify the anisotropy of the fluid. Good agreement is found for moderate anisotropy when compared to associated results from finite-element simulations. At pronounced anisotropy, the induced flow fields are still predicted qualitatively correctly by the perturbative theory, although quantitative deviations arise. We hope to stimulate with our investigations corresponding experimental analyses, for example, concerning fluid flows in anisotropic thin films on uniaxially rubbed supporting substrates.         
	\end{abstract}
	
	\maketitle
	
	
	\section{Introduction}

	The diffusional motion of biological membrane components, including lipids and proteins, plays a crucial role in living systems, governing numerous biochemical functions.
	{Notable examples include signal transduction, where the movement of receptor proteins and their interactions with ligands facilitate the transmission of signals across the cell membrane~\cite{kholodenko2003four, picard2022mitochondrial}, and cellular communication, which relies on the diffusion of molecules like neurotransmitters and hormones across synaptic and cellular membranes~\cite{alberts2002general, choquet2013dynamic}.}
	Early theoretical investigations into the Brownian motion of particles within a membrane date back to Saffman and Delbr\"uck, who determined the hydrodynamic resistance experienced by a slender disk in motion within a thin sheet of viscous fluid representing a model of a lipid bilayer~\cite{saffman1975brownian, saffman1976brownian}
	{In their seminal works, they assumed that the viscosity of the fluid representing the membrane is much greater than that of the surrounding exterior liquid, defining the limit $\operatorname{Bq} \gg 1$ for the Boussinesq number $\operatorname{Bq} = \eta_\mathrm{S} / a\eta$. Here, $\eta_\mathrm{S}$ is the surface viscosity, $\eta$ is the viscosity of the exterior fluid, and $a$ is the radius of the disk.}
	Later Hughes \textit{et al.}~\cite{hughes1981translational} addressed the scenario involving arbitrary viscosities, providing exact expressions for resistance coefficients through techniques of dual integral equations.
	{
		The problem with finite subphase depth was subsequently addressed by Stone and Ajdari~\cite{Stone.Ajdari1998}, partially through numerical methods. For infinite depth and vanishing surface viscosity ($\operatorname{Bq} \to 0$), the translational drag coefficient was found to be 50\% larger than the drag coefficient for a fluid with a free surface~\cite{Stone.Ajdari1998, daddi2024hydrodynamic}. 
		More recently, the translational motion of a disk embedded in a nearly inviscid Langmuir film, marking the limit $\operatorname{Bq} \ll 1$, was revisited~\cite{yariv2023motion}.
		By utilizing the reciprocal theorem~\cite{masoud2019reciprocal} in a fluid domain tailored to the asymptotic topology of the problem under investigation, the leading-order correction to the drag coefficient has been analytically obtained.
	}
	An extension of the Saffman theory was proposed beyond conventional hydrodynamic theories by introducing a generic model involving a protein coupled to a nonconserved scalar order parameter, such as chain stretching.~\cite{camley2012contributions}.
	{In that model, the diffusion coefficient is found to scale inversely with protein radius when the protein is much larger than the order parameter correlation length. This contrasts with the hydrodynamic Saffman-Delbrück theory, which predicts a logarithmic dependence of the diffusion coefficient on protein radius.
		The model is observed to align more closely with experimental findings that have measured protein diffusion coefficients, revealing a stronger correlation with protein radius~\cite{gambin2006lateral, gambin2010variation, kriegsmann2009translational}.}
	On the experimental side, progress in fluorescence imaging techniques enabled a monitoring of the dynamics and interactions of proteins within living cells~\cite{lippincott2001studying, reits2001fixed, capoulade2011quantitative, caneque2018visualizing}.

	Furthermore, the dynamics of particles bound to supported membranes, that is, membranes attached to a substrate, attracted additional attention~\cite{izuyama1988}.
	Besides maintaining the fluid membrane in a flat state, the frictional effect of the rigid substrate with the liquid membrane layer is commonly taken into account by introducing a linear friction term into the hydrodynamic equations. Mathematically, this approach is similar to the formulation initially proposed by Brinkman to describe flows in porous media~\cite{brinkman1949calculation}.
	It offers the advantage of a linear equation that allows for analytical treatment, eliminating the necessity of solving complex dual integral equations encountered in the Saffman-Delbr\"uck model~\cite{saffman1976brownian}.
	Including the frictional term serves to regularize the solution, addressing a well-known issue -- the absence of a steady-state solution for the lateral translation of an infinitely extended, rigid cylinder in the linearized Stokes flow equations under low-Reynolds-number conditions. This problem is often referred to as the Stokes paradox. It maps, in two dimensions, to the lateral translation of a rigid disk in a two-dimensional thin fluid film. If the flow equations are solved applying a net lateral translational force to the cylinder or disk, a logarithmic divergence in the stationary flow equations emerges. 
	Employing a comparable framework that includes momentum decay, the resistance coefficient has further been derived for a circular liquid domain with finite viscosity in motion within a two-dimensional membrane~\cite{ramachandran2010drag}.

	The rheological properties of actual membranes typically deviate from Newtonian behavior.
	In various biologically relevant scenarios, different types of particles 
	bound to membranes often 
	encounter the challenge of navigating through membranes characterized by complex properties~\cite{manikantan2020surfactant}.
	{
		Levine \textit{et al.} extended the classical model
		for extended bodies embedded within infinitely extended viscous films and membranes, addressing both in- and out-of-plane dynamics~\cite{levine2004dynamics, levine2004mobility}.
		It has been found that for large inclusions, both rotation and motion perpendicular to the long axis demonstrate purely local drag behavior, where the drag coefficient is algebraic in terms of the dimensions of the elongated object.
	}
	Moreover, hydrodynamic forces acting on laterally moving inclusions within a two-dimensional liquid domain confined by a supported fluid membrane, considering the concept of ``odd viscosity'', have been explored in various contexts by Hosaka \textit{et al.}~\cite{hosaka2021nonreciprocal, hosaka2021hydrodynamic, hosaka2023hydrodynamics}.
	{The authors report that a disk of finite size moving in the two-dimensional fluid does not only experience a dissipative drag force that opposes its motion, but also a non-dissipative lift force.}
	More recently, the hydrodynamic aggregation of membrane-bound inclusions was examined under the influence of non-Newtonian surface rheology~\cite{vig2023hydrodynamic}.
	{In that case, a corrective force arises due to the interplay between membrane flow and non-constant viscosity. This suggests a mechanism for biologically favorable protein aggregation within membranes.}

	In mixtures with insoluble surfactant species, phase separation is frequently observed. It approximately leads to the formation of two-dimensional dispersions composed of condensed ordered aggregates suspended within an extended disordered fluid phase underneath~\cite{lingwood2010lipid, sezgin2017mystery}.
	While much attention has been devoted to exploring the dynamics in various complex scenarios, to the best of our knowledge, the hydrodynamics of a disk in a thin orientationally ordered, uniaxially anisotropic, two-dimensional fluid film has not been investigated thus far.
	In that context, liquid crystals show phases of matter with physical characteristics between those of regular disordered liquids and translationally and/or orientationally ordered solid crystalline states~\cite{de1993physics}. 
	If forming uniaxially anisotropic -- and in that sense orientationally ordered -- phases, these substances are typically composed of elongated rod-like or flat disk-like organic molecules. The axis of their collective orientational ordering is referred to as the \enquote{director}~\cite{lavrentovich2016active}.
	Rheological properties of such nematic liquid crystals have been studied extensively~\cite{kneppe1981determination, ehrentraut1995viscosity}.

	Hydrodynamic couplings between the nematic director and imposed or resulting velocity fields is frequently described within a framework of continuum mechanics rooted in the Leslie-Ericksen theory~\cite{leslie1968some, ericksen1969continuum, stephen1974physics}. 
	A formulation of the hydrodynamic characteristics of liquid crystals based on a slightly different point of view was suggested by Martin \textit{et al.}~\cite{martin1970new}.
	This alternative approach relies on the fundamental laws of conservation of mass, energy, and momentum. Corresponding equations are coupled to general equations resulting from spontaneously broken symmetries and involving the associated variables~\cite{forster2018hydrodynamic, pleiner1996hydrodynamics}.
	
	{
		The dynamics of colloidal particles immersed in a three-dimensional liquid crystal has been extensively studied over many years~\cite{ruhwandl1995friction, stark99, stark01review, turiv13}. 
		Various types of anchoring and alignment of liquid crystal  molecules on particle surfaces can elastically distort an otherwise spatially uniform alignment of the director field~\cite{cordoba16}. The axisymmetric flow field around a sphere dragged along the director of an undisturbed, aligned nematic liquid crystal  has been theoretically studied, particularly when neglecting one viscosity coefficient, yielding analytical expressions for the frictional drag acting on the sphere in a quiescent liquid crystal~\cite{kneppe91}. 
		Later, a general solution for arbitrary orientation and viscosity coefficients was obtained numerically~\cite{heuer92}. 
		Additionally, closed-form analytical formulas derived from conservation laws for nematic liquid crystals have been developed using techniques of Fourier transformation~\cite{gomez13, gomez-gonzalez16}. Alongside, the Stokes drag of a spherical particle in various nematic environments has been investigated through computer simulations~\cite{stark01, stark02, loudet04, stark02jpcm}.
		Experimental studies have shown that self-diffusion in a nematic liquid crystal follows a generalized Stokes-Einstein relation, with the effective diffusion coefficient along the oriented far-field director typically greater than perpendicular to it under the given circumstances~\cite{pasechnik04, hasnain06, gleeson06, mondiot12}.
	}

	In the present paper, we address the lateral motion of a rigid disk laterally surrounded by a weakly uniaxially orientationally ordered fluid. We confine ourselves to steady-state solutions under overdamped conditions, that is, for low-Reynolds-number flows in a two-dimensional setting. The dynamics involves five coefficients of viscosity. We assume the additional effect of linear damping of flow through linear friction, for instance, with a supporting substrate of the fluid film or a surrounding fluid. Mainly, we concentrate on developing an analytical framework, which is then tested against results from numerical simulations. 

	In this study, we utilize a method of two-dimensional Fourier transformation to examine how a disk behaves within a two-dimensional fluid film of uniaxial anisotropy. Previously, {this method} has been successfully employed to analyze dynamics near various surfaces, including walls and interfaces~\cite{felderhof05, swan07,  felderhof10loss, swan10, daddi18jpcm}, as well as to understand the behavior of elastic membranes undergoing different modes of shear and bending deformation~\cite{bickel06, felderhof06, daddi16, daddi16b, daddi16c, daddi17, daddi18epje, daddi18jcp}. Thus, widely employed in studying elastic sheets and flexible membranes, this technique capitalizes on the symmetry that exists across horizontal planes. The central idea involves setting boundary conditions for the velocity on the perimeter of the disk and there determining the force density as an unknown parameter. 
	{There are well-established approaches available in classical textbooks on fluid mechanics that provide analogous methods~\cite{happel12, kim13, dhont96}. However, they are primarily applicable to isotropic fluids.}

	We demonstrate that under isotropic additional linear friction, for instance, with a supporting substrate or fluid, the drag force experienced by a translating disk in an anisotropic two-dimensional fluid medium remains independent of the direction of motion to the considered order in anisotropy of the two-dimensional fluid. This observation stands in strong contrast to the three-dimensional scenario, where the drag coefficient varies with the direction of motion \cite{gomez13}. Furthermore, we rigorously evaluate the accuracy and validity of the approximate perturbative expansion utilized in our study through close comparison with finite-element simulations. 

	The remainder of the presentation is organized as follows.
	In Sec.~\ref{sec:mathematischeFormulierung}, we provide an overview of the continuum description at low Reynolds numbers, employed to characterize the dynamics of the surrounding two-dimensional anisotropic sheet of ordered nematic fluid.
	In this context, we utilize Fourier transformations for the hydrodynamic fields.
	From the inverse Fourier transformation, we obtain in Sec.~\ref{sec:hydrodyn} perturbative analytical solutions for the flow field induced by a disk laterally translating in a weakly nematic fluid sheet.
	Next, in Sec.~\ref{sec:resistance}, we evaluate the impact of fluid anisotropy on the translational resistance coefficients of the disk laterally surrounded by the nematic fluid. Additionally, we present asymptotic expressions of the resistance coefficients for the limiting cases of weak and strong linear friction with a supporting substrate or other types of surroundings.
	In Sec.~\ref{sec:FEM}, we compare our theoretical predictions with finite-element simulations. Following this comparison, an analysis of the results is provided in Sec.~\ref{sec:res_discussion}. Conclusions are added in Sec.~\ref{sec:conclusions}.
	{Besides, Appendix~\ref{appendix:tab} contains a summary of the key parameters introduced in this manuscript for convenient reference. Appendix~\ref{appendix:int} presents details of calculating a non-trivial infinite integral in Sec.~\ref{sec:hydrodyn}.}

	\section{Low-Reynolds-number flows in a two-dimensional anisotropic fluid medium}
	\label{sec:mathematischeFormulierung}
	
	\subsection{Generalized Stokes hydrodynamics}
	
	{For flows at low Reynolds numbers, viscous contributions dominate inertial ones.} Under these conditions, the dynamics of the surrounding fluid are predominantly governed by the Stokes equation. We here study an extension of this equation to take into account possible friction with the surrounding. In reality, such friction may arise from a substrate and possibly by an additional cover slide confining a thin, basically two-dimensional fluid layer, or, possibly and to some approximation, from the friction with a surrounding fluid. Consequently, the flow equation can be generally expressed as
	\begin{equation}
		\nabla_j\sigma_{ij} (\R) + M_{ij} v_j (\R) = f_i(\bm{r})  \, . \label{eq:Stokes}
	\end{equation}
	In this equation, $\sigma_{ij}(\R) = p(\R) \, \delta_{ij} + \widetilde{\sigma}_{ij} (\R)$ represents the components of the stress tensor, where $p(\R)$ is the pressure field, $\delta_{ij}$ is the Kronecker delta, $f_i(\bm{r})$ in two dimensions signifies a force density acting on the fluid, and summation over repeated indices is implied following Einstein's notation. 
	$M_{ij}$ represent the coefficients of linear friction with the substrate or top and bottom surroundings. $\vect{v} (\R)$ is the velocity field. (In a different context, $\bm{M}$ played the role of a permeability tensor of SI unit kg/s$\,$m$^2$,  particularly when considering flows through porous media \cite{kohr08}.)

	For a simple isotropic fluid, $\bm{\widetilde{\sigma}} (\R)$ introduces viscous dissipation into the dynamic equations, encompassing two viscosity parameters --- one associated with shear viscosity and the other with dissipation during volume changes.
	Instead, we here address a uniaxially symmetric, anisotropic fluid medium. 
	We start from the corresponding symmetry-based expressions for a conventional nematic liquid crystal deep within the nematic phase~\cite{martin72,forster75book}. 
	The local orientation of the axis of uniaxial order is defined by the normalized director field $\bm{\hat{n}}(\bm{r})$. 
	
	Generally, the director field is dynamic, changing over time, for example, due to fluid flows or the action of external fields. Spatial variations in $\bm{\hat{n}}(\bm{r})$ require deformational energy and contribute to the stress tensor~$\bm{\widetilde{\sigma}} (\R)$. 

	In our considerations, we examine a scenario of the director being consistently and uniformly oriented along a single global axis \cite{kneppe81,heuer92,kos18}. This alignment can be achieved, for instance, through the influence of a strong aligning external electric field (under insulating conditions) or a magnetic field \cite{degennes95,pleiner96,pleiner02}. When $\bm{\hat{n}} (\R)$ is perfectly aligned with the field, there are not any associated gradients contributing to the stress tensor.

	Hence, the stress tensor $\bm{\widetilde{\sigma}} (\R)$ is simplified to the pure dissipative stress resulting from gradients in the fluid flow~\cite{pleiner96,pleiner02},
	\begin{equation}
		\widetilde{\sigma}_{ij} = \sigma_{ij}^{\text{D}} = {}-\nu_{ijkl}\nabla_l v_k \, . \label{eq:momentumEquation}
	\end{equation}
	In this expression, $\nu_{ijkl}$ refer to the components of the surface viscosity tensor of uniaxial symmetry of SI unit kg/s, given by \cite{pleiner96}
	\begin{align}
		\nu_{ijkl} &= \nu_2(\delta_{ik}\delta_{jl} + \delta_{il}\delta_{jk})
		+ \bar{\nu} n_in_jn_kn_l + (\nu_4-\nu_2)\delta_{ij}\delta_{kl} \notag \\[3pt]
		&\quad+ (\nu_3-\nu_2)(n_in_k\delta_{jl}+n_in_l\delta_{jk}+n_jn_k\delta_{il}+n_jn_l\delta_{ik}) \notag \\[3pt]
		&\quad+ (\nu_5-\nu_4+\nu_2)(\delta_{ij}n_kn_l+\delta_{kl}n_in_j) \, , \label{eq:nu_ijkl}
	\end{align}
	where we introduced the abbreviation 
	\begin{equation}
		\bar{\nu} = 2 \left( \nu_1 + \nu_2 - 2\nu_3 \right) \, ,
	\end{equation}
	following Ref.~\onlinecite{daddi2018dynamics}.

	From now on, we do not explicitly mention the dependence of the fields on $\R$ any longer.
	Overall, we incorporate the common assumption made in similar analyses of fluid flows, namely, local volume conservation and constant density of the fluid. As a result, the continuity equation simplifies to
	\begin{equation}
		\boldsymbol{\nabla} \cdot\vect{v}=0 \, . \label{eq:incompressibilityEqn}
	\end{equation}

	Under the given circumstances, the director $\bm{\hat{n}}$ introduces uniaxial anisotropy to the viscosity tensor.
	Without loss of generality, we choose $\bm{\hat{n}} \parallel \bm{\hat{x}}$. Thus, the components of the viscous stress tensor in Eq.~(\ref{eq:nu_ijkl}) in Cartesian coordinates become
	\begin{align}
		-\tilde{\sigma}_{ij} &= 
		\left( 2\nu_1+\nu_2-\nu_4+\nu_5 \right)  v_{x,x} \, \delta_{ix} \delta_{jx} \notag \\[3pt]
		&\quad+ \left( \nu_2+\nu_4 -\nu_5 \right) v_{y,y} \, \delta_{iy} \delta_{jy} \notag \\[3pt]
		&\quad+ \nu_3 \left( v_{x,y}+v_{y,x} \right) \left( 
		\delta_{ix}\delta_{jy}+\delta_{iy}\delta_{jx}\right). 
		\label{eq:sigmaD}
	\end{align}
	Commas denote partial derivatives.

	In the subsequent discussion, {we 
		examine the scenario of one of the principal axes of the friction tensor aligning with the nematic director. Thus, the friction tensor in our representation adopts a diagonal form}
	\begin{equation}
		\bm{M} = \left( 
		\begin{matrix}
			m_\parallel^2 & 0 \\
			0 & m_\perp^2 
		\end{matrix}
		\right) . \label{eq:friction}
	\end{equation}

	Inserting Eq.~\eqref{eq:sigmaD} into Eq.~\eqref{eq:Stokes} and subtracting $\left( \nu_3 + \nu_5 \right) \boldsymbol{\nabla} \left( \boldsymbol{\nabla} \cdot \vect{v} \right)=0$ from the resulting equations, we obtain
	\begin{subequations}
		\begin{align}
			p_{,x} -\nu_3 \left( \zeta_1 v_{x,xx} + v_{x,yy} - \alpha_\parallel^2 v_x \right)  &= f_x \, , \\[3pt]
			p_{,y} -\nu_3 \left( v_{y,xx} + \zeta_2 v_{y,yy} - \alpha_\perp^2 v_y \right) &= f_y \, .
		\end{align}
		\label{eq:gov_eqns_final}
	\end{subequations}
	We here defined the dimensionless viscosities $\zeta_1 = \left( 2\nu_1+\nu_2-\nu_4+\nu_5 \right)/\nu_3 - 1$ and $\zeta_2 = \left( \nu_2+\nu_4-\nu_5\right)/\nu_3 - 1$.
	Furthermore, for the sake of convenience, we introduce the abbreviations $\alpha_\parallel^2 = m_\parallel^2 / \nu_3$ and $\alpha_\perp^2 = m_\perp^2 / \nu_3$, both possessing dimensions of inverse length squared.
	Notably, we recover the classical Brinkman equation~\cite{brinkman1949calculation} for the isotropic case, if $\nu_1=\nu_2=\nu_3=\nu_4$ and $\nu_5 = 0$.

	\subsection{Fourier representation}

	To obtain the solution for the flow velocity and pressure fields, a two-dimensional Fourier transformation along both the $x$ and $y$ directions is employed. 
	We define the forward Fourier transformation of a given function $g(\boldsymbol{\rho})$ as
	\begin{equation}
		\widetilde{g}(\bm{k}) = \mathscr{F}\left\{ g(\boldsymbol{\rho}) \right\} = \int_{\mathbb{R}^2} g(\boldsymbol{\rho}) \, e^{-i \bm{k} \cdot \boldsymbol{\rho}} \, \mathrm{d}^2 \boldsymbol{\rho} \, ,
	\end{equation}
	where $\bm{k}$ represents the wavevector, and its inverse as
	\begin{equation}
		g(\boldsymbol{\rho}) = \mathscr{F}^{-1}\left\{ \widetilde{g}(\bm{k}) \right\} = 
		\frac{1}{\left(2\pi\right)^{2}} \int_{\mathbb{R}^2} \widetilde{g}(\bm{k}) \, e^{i \bm{k} \cdot \boldsymbol{\rho}} \, \mathrm{d}^2 \bm{k} \, .
	\end{equation}
	In these expressions, $\boldsymbol{\rho} = (x, y)$ denotes the position vector. 
	Furthermore, we introduce the wavenumber $k = |\bm{k}|$ and the unit wavevector $\bm{\hat{k}} = \bm{k}/k$.
	We employ a polar coordinate system to express $\bm{\hat{k}}=(\hat{k}_x , \hat{k}_y)= (\cos\phi, \sin\phi)$.
	
	Next, we define the signed dimensionless number
	\begin{equation}
		A = \frac{\bar{\nu}}{\nu_3} \, . \label{eq:A}
	\end{equation}
	It is related to the degree of anisotropy of the uniaxial, nematic fluid. $A=0$ in the isotropic case.
	
	{
		For the common representatives of liquid crystals MBBA (N-(4-methoxybenzylidene)-4-butylaniline) and 5CB (4-cyano-4'-pentylbiphenyl), we estimate a typical value of \(A \approx 4\) based on literature data~\cite{stewart04, daddi2018dynamics}.
		Physically, it is important to highlight that A must meet the condition \(A > -4\), which is crucial for thermodynamic stability. This condition is derived from the necessity that \(A > \nu_2/\nu_3 - 4\). 
		Given that both \(\nu_2\) and \(\nu_3\) must be positive, it follows that \(A > -4\).
		Below, we demonstrate how this parameter serves as a perturbation parameter in obtaining approximate solutions for the hydrodynamic flow fields.
	}
	
	{
		The solution for the velocity field in Fourier space can be obtained by taking the Fourier transformations of Eqs.~\eqref{eq:gov_eqns_final}. Subsequently, we apply the divergence-free condition of the velocity field to eliminate the pressure field from these equations.}
	Introducing the Green's function $\widetilde{\bm{\mathcal{G}}} (k, \phi)$, the fluid velocity in  Fourier space can be expressed as
	\begin{equation}
		\widetilde{\vect{v}} (k, \phi) = \widetilde{\bm{\mathcal{G}}} (k, \phi) \cdot \widetilde{\bm{f}} (k, \phi) \, . 
		\label{eq:velo_fourier}
	\end{equation}
	Here, 
	\begin{equation}
		\widetilde{\bm{\mathcal{G}}} (k, \phi) =
		\frac{1}{ \nu_3 H (k, \phi) } \,
		\bm{G} (\phi) \, , \label{eq:tG}
	\end{equation}
	where we defined
	\begin{equation}
		\bm{G} (\phi) = \left(
		\begin{matrix}
			\sin^2 \phi & -\sin\phi \cos\phi \\
			-\sin\phi \cos\phi & \cos^2\phi
		\end{matrix}
		\right) , \label{eq:G}
	\end{equation}
	together with the abbreviations $H (k, \phi) = \alpha_\parallel^2 \sin^2\phi + \alpha_\perp^2 \cos^2\phi + k^2 B(\phi)$ and $B(\phi) = 1 + A \sin^2\phi \cos^2\phi$.
	In the following, we confine our analysis to the case of $A \ge -4$, which ensures $B(\phi) \ge 0$. This condition is essential to guarantee $H(k,\phi) > 0$, a prerequisite for the Green's function to remain well-defined across all values of $k$.
	
	The pressure field in Fourier space is denoted as
	\begin{equation}
		\widetilde{p} (k, \phi) = \widetilde{\bm{\mathcal{P}}} (k, \phi) \cdot \widetilde{\bm{f}} (k, \phi) \, , \label{eq:pressure-fourier}
	\end{equation}
	where
	
	\begin{subequations}
		\begin{align}
			\mathcal{P}_x &= -\frac{i\cos\phi}{kH (k,\phi)} 
			\Big( k^2 \left( \zeta_2 \sin^2\phi + \cos^2\phi \right) + \alpha_\perp^2 \Big) \, , \\[3pt]
			\mathcal{P}_y &= -\frac{i\sin\phi}{kH (k,\phi)} 
			\Big( k^2 \left( \zeta_1 \cos^2\phi + \sin^2\phi \right) + \alpha_\parallel^2 \Big) \, .
		\end{align}
	\end{subequations}

	Next, we derive the corresponding expressions in real space. This includes the definition of the appropriate boundary conditions.

	\section{Hydrodynamic flow fields}
	\label{sec:hydrodyn}

	\begin{figure}
		\centering
		\includegraphics[scale=.6]{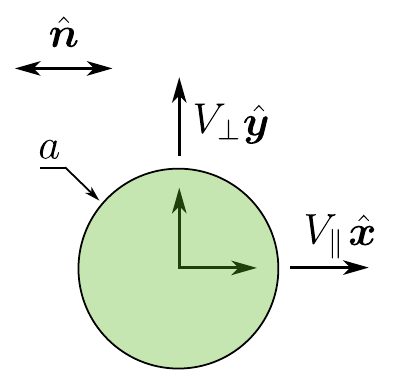}
		\caption{
			{Schematic illustration of the quantities involved for the two-dimensional disk under consideration. The circular disk of radius \(a\) moves in a two-dimensional anisotropic sheet of a globally aligned uniaxial fluid. The axis of anisotropy, characterized by the nematic director \(\hat{\bm{n}}\), is aligned permanently along the \(x\)-direction. Our disk moves 
				with velocity components \(V_\parallel\) parallel and \(V_\perp\) perpendicular to the director.
				We work within the frame of reference anchored to the disk, so that the origin of the coordinate frame remains at the center of the disk.
				The frictional effect with a supporting substrate or top and bottom surroundings is represented by a linear friction tensor.
			}
		}
		\label{fig:disk}
	\end{figure}
	
	We consider the steady motion of a circular disk moving in a two-dimensional anisotropic sheet of a globally aligned uniaxial fluid, specifically, a nematic liquid crystal.
	The disk of radius~$a$ 
	moves with speed~$V$. No-slip boundary conditions for the flow field $\vect(\rho,\theta)$ of the fluid apply at the perimeter of the disk. They ensure $\vect{v}(\rho = a, \theta) = V_\parallel \bm{\hat{x}} + V_\perp \bm{\hat{y}}$, that is, the fluid acquires identical flow velocity over the complete perimeter of the disk. 
	{Our approach of solution concentrates on the comoving frame of reference that permanently anchors its origin to the center of the disk.}
	Leveraging the linearity of the governing equations of motion, we can independently analyze the motion parallel and perpendicular to the nematic director $\bm{\hat{n}} \parallel \bm{\hat{x}}$.
	{Figure~\ref{fig:disk} provides an illustration of the quantities involved in the geometry.}
	{It is important to highlight that our attention is primarily placed on the dynamics of the considered two-dimensional fluid. We do not describe the response of a possible environment surrounding this thin fluid film, for instance, a supporting substrate or a fluid or gaseous component beneath or above the fluid film and the disk. The influence of the surrounding environment, the dynamics of which we do not include in our consideration, is represented by the friction tensor specified by Eq.~\eqref{eq:friction}.
	}

	Our focus now is on the hydrodynamic velocity field around the disk.
	We express the a~priori unknown force distribution that the disk exerts on the fluid across its boundary in the form of a Fourier series,
	\begin{equation}
		\bm{f} (\rho, \theta) = \sum_{n=-\infty}^\infty \delta(\rho-a) \, \bm{f}_n  \, e^{in\theta} \, . 
		\label{eq:force-distr}
	\end{equation}
	The coefficients $\bm{f}_n$ are to be determined based on the given boundary condition. 
	{We note that the dimension of the coefficients $\bm{f}_n$ is not identical to those of the force density $\bm{f}$, because the dimension of the delta function is of inverse length.}
	The overall force exerted by the disk is  derived later through integration of $\bm{f} (\rho, \theta)$ over the area of the fluid, that is, over the area of $\rho\geq a$. Evidently, it is found solely from the zeroth-order Fourier mode, $\langle \bm{F} \rangle = 2\pi a \, \bm{f}_0$.
	Following a straightforward algebraic manipulation, the force density in Eq.~\eqref{eq:force-distr} can be expressed in Fourier space as
	\begin{equation}
		\widetilde{\bm{f}} (k, \phi) = 2\pi a \sum_{n=-\infty}^\infty i^{-n} \, \bm{f}_n J_n(ka) \, e^{in\phi} \, ,
		\label{eq:force_fourier}
	\end{equation}
	where $J_n$ denotes the Bessel function of the first kind of $n$th degree~\cite{abramowitz72}.

	\subsection{Velocity field}
	
	Our central step is to determine the force density distribution around the perimeter of the disk that there satisfies the boundary conditions for the velocity field. 
	Since the governing equations of motion exhibit a linear nature, it is adequate to focus on the predominant two modes of motion, that is, parallel and perpendicular to the director.
	Here, we impose a no-slip boundary condition for the velocity field at the perimeter of the disk. Thus, for the velocity $\bm{V}=V\bm{\hat{V}}$ of the disk, 
	the components of the velocity field {of the fluid} at $\rho = a$ are given by 
	{$v_x(\rho=a) = V$ and $v_y(\rho=a) = 0$}
	for motion parallel to the director ($\bm{\hat{V}}\|\bm{\hat{x}}$). 
	Similarly, they are given by  
	{$v_x(\rho=a) = 0$ and $v_y(\rho=a) = V$}
	for motion perpendicular to the director ($\bm{\hat{V}}\|\bm{\hat{y}}$).
	{By matching the Fourier modes at the surface of the disk, we obtain the solution for the force density in terms of a Fourier series.}
	
	{It is important to clarify that we project vector quantities onto Cartesian coordinates, in accordance with the geometry of the system. Conversely, we express positions both in real and Fourier spaces using polar coordinates.
		By applying inverse Fourier transformation,} the velocity field can be expressed in real space as~\cite{baddour2011two}
	\begin{equation}
		\vect{v}(\rho, \theta) = \left( 2\pi \right)^{-2} 
		\sum_{m=-\infty}^\infty {i^m} \mathbfcal{S}_{m} (\rho) \, {e^{im\theta}} \, , \label{eq:v_forReview}
	\end{equation}
	wherein~$\mathbfcal{S}_{m}$ is a radial function defined by the double integral over the wavenumber~$k$ and angle~$\phi$ as
	\begin{equation}
		\mathbfcal{S}_{m} (\rho) = \int_0^{2\pi} \int_0^\infty \widetilde{\vect{v}} (k,\phi) \,
		{J_{m} (k\rho) \, e^{-im\phi}} \, k \, \mathrm{d}k \, \mathrm{d}\phi \, .
		\label{eq:Sm}
	\end{equation}
	
	{
		Since the boundary conditions prescribed at the surface of the disk correspond to constant velocities, the velocity field given by Eq.~\eqref{eq:v_forReview}, when evaluated at \(\rho = a\), is
		\begin{equation}
			\mathbfcal{S}_m(\rho = a) = \left( 2\pi \right)^{2}\bm{V} \, \delta_{m\, 0} \, ,
			\label{eq:Sm_rho}
		\end{equation}
		for $m \ge 0$.
		Accordingly, only the zeroth mode matches the velocity of the disk at $\rho = a$, while all other modes are set to zero.
	}
	
	{
		Subsequently, utilizing Eqs.~\eqref{eq:velo_fourier} and \eqref{eq:force_fourier}, we find that the $\phi$-integration 
		in 
		Eq.~\eqref{eq:Sm} vanishes, if $m-n$ is an odd number. Since Eqs.~\eqref{eq:Sm_rho} represent a system of linear equations for the unknown values of $\bm{f}_n$, it follows that all odd Fourier modes in the expression of the force density must vanish. Consequently, only the even modes will persist in the series expansion of the velocity field as well.
	}

	Then, the velocity field can be expressed as an infinite double sum of the form,
	\begin{equation}
		\vect{v}(\rho, \theta) = \frac{a}{2\pi\nu_3} 
		\sum_{m=-\infty}^\infty e^{2im\theta} \sum_{n = -\infty}^\infty \, \mathbfcal{Q}_{mn} (\rho) , 
	\end{equation}
	where
	\begin{equation}
		\mathbfcal{Q}_{mn} (\rho) = \int_0^{2\pi}
		\Phi_{mn} (\rho, \phi) \, e^{2i (n-m) \phi}  \, \bm{G} (\phi) \cdot \bm{f}_{2n} \, \mathrm{d} \phi \, ,
		\label{eq:Q}
	\end{equation}
	and 
	\begin{equation}
		\Phi_{mn} (\rho, \phi) = \frac{(-1)^{m+n}}{B}  \int_0^\infty \frac{u \, J_{2m}(u) J_{2n}({w} u)}{u^2 + \sigma^2} \,  \, \mathrm{d} u \, , 
		\label{eq:Phi}
	\end{equation}
	with the dimensionless numbers ${w} = a/\rho$ and
	{
		\begin{equation}
			\sigma  = \frac{ \rho}{B^{1/2}}
			\left( \alpha_\parallel^2 \sin^2\phi + \alpha_\perp^2 \cos^2\phi \right)^{1/2} .
		\end{equation}
		We note that $B$ is a function of $\phi$, as previously defined below Eq.~\eqref{eq:G}.}
	Moreover, we introduced the change of variables $u = k\rho$.
	${w} \le 1$ marks the fluid region outside the disk.
	We recall that the index $m$ is employed in the context of the inverse Fourier transformation of the velocity field, whereas the index $n$ corresponds to the Fourier representation of the unknown force density.

	In the following, we set for motion parallel to the director $\bm{f}_{2n}^\parallel = \left( c_{2n}^\parallel, -id_{2n}^\parallel \right)^\top$, with $(\cdot)^\top$ denoting the transpose operator.
	It follows that $c_{-2n}^\parallel = c_{2n}^\parallel$ and $d_{-2n}^\parallel = -d_{2n}^\parallel$, implying that $d_0^\parallel = 0$.
	For motion perpendicular to the director, we set $\bm{f}_{2n}^\perp = \left( -ic_{2n}^\perp, d_{2n}^\perp \right)^\top$ so that $c_{-2n}^\perp = -c_{2n}^\perp$ and $d_{-2n}^\perp = d_{2n}^\perp$, implying that $c_0^\perp = 0$.
	These coefficients are selected as real numbers. Their values are determined from the specified boundary conditions, as detailed below.

	In this way, the Fourier representation of the projected force as given by Eq.~\eqref{eq:force_fourier} can be expressed as
	\begin{equation}
		\widetilde{\bm{f}}_{\parallel} = 
		\sum_{n=0}^\infty s_{2n} \left( c_{2n}^\parallel \cos (2n\phi) \, \bm{\hat{x}} + d_{2n}^\parallel \sin (2n\phi) \, \bm{\hat{y}} \right)
	\end{equation}
	for motion parallel to the director, and
	\begin{equation}
		\widetilde{\bm{f}}_{\perp} = 
		\sum_{n=0}^\infty s_{2n} \left( c_{2n}^\perp \sin (2n\phi) \, \bm{\hat{x}} + d_{2n}^\perp \cos (2n\phi) \, \bm{\hat{y}} \right) 
	\end{equation}
	for motion perpendicular to the director, with 
	\begin{equation}
		s_{2n} = 4\pi a (-1)^n \left( 1 -\frac{ \delta_{n\, 0} }{2} \right)
		J_{2n} (ka) \, .
	\end{equation}

	Accordingly, the velocity field can be written as
	\begin{align}
		\vect{v}{^k} (\rho, \theta) &=  \frac{a}{\nu_3} 
		\sum_{m = 0}^\infty \sum_{n = 0}^\infty
		\left( { U_{mn}^k \, \bm{\hat{x}} + V_{mn}^k  \, \bm{\hat{y}} } \right) , 
	\end{align}
	{for $k \in \{ \parallel, \perp\}$,} wherein
	\begin{subequations} \label{eq:UV}
		\begin{align}
			{U_{mn}^k} &= \phantom{+} \frac{2}{\pi} \int_0^{2\pi} 
			{\mathcal{M}_{mn}^k} (\theta, \phi) \, \Phi_{mn} (\rho, \phi) \, \sin \phi \, \mathrm{d}\phi \, , \\[3pt]
			{V_{mn}^k} &= -\frac{2}{\pi} \int_0^{2\pi} 
			{\mathcal{M}_{mn}^k} (\theta, \phi) \, \Phi_{mn} (\rho, \phi) \, \cos \phi \, \mathrm{d}\phi \, ,
		\end{align}
	\end{subequations}
	with
	\begin{equation}
		{\mathcal{M}_{mn}^k} (\theta, \phi) = \xi_{mn}
		\cos \left( 2m(\phi-\theta) \right)
		{Z_{2n}^k (\phi)} \, ,
	\end{equation}
	and 
	\begin{equation}
		\xi_{mn} = \left( 1 -\frac{ \delta_{m\, 0} }{2} \right) \left( 1-\frac{ \delta_{0\, n} }{2} \right) \, , 
	\end{equation}
	so that $\xi_{mn} = 1/4$ if $m=n=0$, $\xi_{mn} =1/2$ if $m=0$ or $n=0$, and $\xi_{mn} =1$ otherwise.
	In addition,
	\begin{equation}
		Z_{2n}^\parallel (\phi) = c_{2n}^\parallel \cos (2n\phi) \sin\phi - d_{2n}^\parallel \sin (2n\phi) \cos\phi \,
	\end{equation}
	for motion parallel to the director, and
	\begin{equation}
		Z_{2n}^\perp (\phi) = d_{2n}^\perp \cos (2n\phi) \cos\phi -c_{2n}^\perp \sin (2n\phi) \sin\phi  \, 
	\end{equation}
	for motion perpendicular to the director.

	As evident from Eqs.~\eqref{eq:Phi} and \eqref{eq:UV}, determining the velocity field involves the assessment of a double integral. Due to the delicate nature of analytical evaluation, especially in the general case, we adopt the following approach.
	The integral over the scaled wavenumber, as defined by Eq.~\eqref{eq:Phi}, allows for exact evaluation, yielding an analytical expression. However, when calculating the ensuing integrals over $\phi$ outlined in Eqs.~\eqref{eq:UV}, we resort to approximations.

	Defining
	\begin{equation}
		\Psi_{mn} = I_{2n}({w}\sigma) \, K_{2m}(\sigma) \, , 
		\label{eq:Psi}
	\end{equation}
	it follows that
	\begin{equation}
		\Phi_{mn} = \frac{1}{B} 
		\begin{cases}
			\Psi_{mn} & \text{ if}\quad m \le n \, , \\
			\Psi_{mn} - \mathcal{P} \Psi_{mn} & \text{ if}\quad m > n \, ,
		\end{cases} \label{eq:improper_int}
	\end{equation}
	where $I_{2n}$ and~$K_{2n}$ represent the modified Bessel functions (also known as the hyperbolic Bessel functions) of the first and second kind, respectively.
	Denoted as $\mathcal{P} \Psi_{mn}$, the principal part of the series expansion of $\Psi_{mn}$ around $\sigma = 0$ refers to its component involving negative powers~\cite{carrier2005functions}. 
	To clarify, the principal part corresponds to the segment of the Laurent series expansion of $\Psi_{mn}$ characterized by negative exponents.
	This result only holds for ${w} \le 1$, which corresponds to the area outside the disk.
	{A detailed analytical solution is provided in Appendix~\ref{appendix:int}.}

	In Tab.~\ref{tab:PP}, we provide the corresponding values of $\mathcal{P} \Psi_{n+1,n}$, $\mathcal{P} \Psi_{n+2,n}$, and $\mathcal{P} \Psi_{n+3,n}$ for $n = 0, \dots, 4$.
	The expressions for $\mathcal{P} \Psi_{mn}$ where $m>n$ can be readily derived for any given values of $m$ and $n$ using computer algebra systems such as Maple~23 that we used in the present work~\cite{maple23}. 
	This can be achieved by calculating the Laurent series expansion of $\Psi_{mn}$ and considering exclusively the terms of negative exponents.
	
	\begin{table*}
		\centering
		{\renewcommand{\arraystretch}{2.}
			\begin{tabular}{|c|c|c|c|}
				\hline
				~$n$~ & $m=n+1$ & $m=n+2$ & $m=n+3$ \\
				\hline
				0 & $2t$ & $-4t \left( 1-3{w}^2 - 12t \right)$ & 
				$6t \big[ 1-8{w}^2+10{w}^4   - 32  \left( 1-5{w}^2 \right)t + 640 t^2 \big] $ \\
				\arrayrulecolor{lightgray}\hline
				1 & $6{w}^2 t$ & $-8{w}^2 t \left( 3-5{w}^2-60t \right)$ & 
				$10 {w}^2 t \big[ 6-24{w}^2+21{w}^4 - 96  \left( 3-7{w}^2 \right) t + 8064 t^2 \big]$ \\
				\hline
				2 & $10{w}^4 t$ & $-12{w}^4 t \left( 5-7{w}^2-140t \right) $ & 
				$14{w}^4 t \big[  15-48{w}^2+36{w}^4 - 192 \left( 5-9{w}^2 \right) t + 34560 t^2 \big]$ \\
				\hline
				3 & $14{w}^6 t$ & $-16{w}^6 t  \left( 7-9{w}^2 - 252 t \right)$ & 
				$18 {w}^6 t \big[ 28-80{w}^2+55{w}^4  - 320  \left( 7-11{w}^2 \right) t + 98560 t^2 \big]$
				\\
				\hline
				4 & $18{w}^8 t$ & ~~$-20 {w}^8 t  \left( 9-11{w}^2 - 396t \right)$~~ & 
				~~$22 {w}^8 t \big[ 45-120{w}^2+78{w}^4 - 480 \left( 9-13{w}^2 \right) t + 224640 t^2 \big]$~~ \\
				\hline
				\vdots & \vdots & \vdots & \vdots \\
				\hline
				$n$ & $\mathcal{P} \Psi_{n+1, n}$ & $\mathcal{P} \Psi_{n+2, n}$ & $\mathcal{P} \Psi_{n+3, n}$ \\
				\arrayrulecolor{black}\hline
			\end{tabular}
		}
		\caption{
			The principal part of the series expansion of $\Psi_{mn}$ is examined for cases of $m=n+1$, $m=n+2$, and $m=n+3$, with $\Psi_{mn}$ defined by Eq.~\eqref{eq:Psi}. General expressions applicable to values of $n$ are provided by Eqs.~\eqref{eq:PP3}.
			{Here, $t=1/\sigma^2$.}
		}
		\label{tab:PP}
	\end{table*}
	
	{Defining $t=1/\sigma^2$,} the first terms of $\mathcal{P} \Psi_{mn}$ for $m=n+1$, $m=n+2$, and $m=n+3$ are obtained as
	\begin{widetext}
		\begin{subequations} \label{eq:PP3}
			\begin{align}
				\mathcal{P} \Psi_{n+1, n} &= 2(2n+1) {w}^{2n} t \, , \\
				\mathcal{P} \Psi_{n+2, n} &= -4(n+1) {w}^{2n} t \left[ 2n+1-(2n+3){w}^2 - 4(2n+1)(2n+3) t \right] \, , \\
				\mathcal{P} \Psi_{n+3, n} &=2  (2n+3) {w}^{2n} t
				\big[ 
				(n+1)(2n+1)-4(n+1)(n+2){w}^2+(n+2)(2n+5){w}^4 \notag \\
				&\quad+16(n+1)(n+2) \left( (2n+5){w}^2-(2n+1) + 4(2n+1)(2n+5)t \right) t
				\big].
			\end{align}
		\end{subequations}    
	\end{widetext}

	We now {define the dimensionless number}
	\begin{equation}
		\beta = 1 - \frac{\alpha_\perp}{\alpha_\parallel} \, , \label{eq:beta_def}
	\end{equation}
	which quantifies the degree of anisotropy in the friction tensor.
	
	As previously noted, the evaluation of integrals in Eqs.~\eqref{eq:UV} is complex. 
	To facilitate analytical advances, we investigate the scenario of weak fluid anisotropy, that is, $|\beta| \ll 1$. 
	Furthermore, we restrict our analysis to cases of $|A| \ll 1$, 
	see Eq.~\eqref{eq:A}.
	Both $A$ and $\beta$ are signed quantities, allowing for both positive and negative values in a physical context.
	Hence, the integrands presented in Eqs.~\eqref{eq:UV} can be expressed as a multi-Taylor expansion in the small parameters $A$ and~$\beta$.
	This facilitates a straightforward analytical evaluation of the resulting integrals with respect to $\phi$ in Eqs.~\eqref{eq:UV}, which otherwise is not obvious.
	Our considerations are limited to terms up to the first order in the multi-Taylor series expansions.

	To determine the unknown series coefficients~$c_{2n}^\parallel$ and~$d_{2n}^\parallel$ for motion parallel to the director, as well as~$c_{2n}^\perp$ and~$d_{2n}^\perp$ for motion perpendicular to the director, we apply the boundary conditions for the velocity field at the surface of the disk. This leads us to a linear system of equations to be solved.
	We find that, to leading orders in $A$ and $\beta$, only the first, third, and fifth Fourier modes in the radial and azimuthal velocities persist without vanishing. 
	The seventh Fourier mode for pressure also persists without vanishing.
	By accounting for the second-order terms, we find that the seventh and ninth Fourier modes persist for the velocity in addition to the eleventh Fourier mode for the pressure.

	To enhance clarity in elucidating the resulting complex mathematical expressions, we introduce the dimensionless quantities $r = \alpha_\parallel \rho$ and ${c} = \alpha_\parallel a$, representing the rescaled radial distance and the rescaled disk radius, respectively.
	%
	%
	The final expressions for the radial and azimuthal components of the fluid velocity field can be cast in the form
	\begin{equation}
		\vect{v} = V \left( \vect{v}^\mathrm{IB} + \vect{v}^* \right) \, . 
	\end{equation}
	In this context, $\vect{v}^\mathrm{IB}$ represents the solution corresponding to an \textit{isotropic Brinkman} fluid of $A=\beta=0$, while~$\vect{v}^*$ denotes the leading-order correction arising from anisotropy.
	%
	%
	The former is obtained as~\cite{martin2019two}
	\begin{subequations}
		\begin{align}
			v_\rho^\mathrm{IB} &= \frac{{c}^2 q_3 - 2r q_5}{r^2 q_1} \, \cos\theta \, , \\
			v_\theta^\mathrm{IB} &= 
			\frac{{c}^2 q_3 - r^2 \left( q_4+q_6 \right)}{r^2 q_1}
			\, \sin\theta \, ,
		\end{align}
	\end{subequations}
	with the abbreviation
	\begin{equation}
		\vect{q} = \big( K_0 ( {c} ), K_1 ( {c} ), K_2 ( {c} ), K_0 ( r ), K_1 ( r ), K_2 ( r ) \big)^\top \, .
	\end{equation}

	We find that the radial and azimuthal components of the leading-order correction to the velocity field involve only terms of the form $\cos(n\theta)$ and $\sin(n\theta)$, respectively, with $n \in \{1,3,5\}$.
	Due to their complexity and lengthiness, the corresponding expressions are not explicitly shown here.
	Nevertheless, the Maple script providing the final expressions of the velocity field is available through the Zenodo repository.

	\subsection{Pressure field}
	
	To complete the picture, we determine expressions for the corresponding pressure field. 
	Results are also derived to leading order in the dimensionless numbers $A$ and $\beta$, incorporating the viscosity ratio
	\begin{equation}
		{b} = \frac{\nu_1 - \nu_4 + \nu_5}{\nu_3} \, .
	\end{equation}
	In the present framework there are no restrictions on the magnitude {or sign} of~$b$. It can be arbitrarily small or large.
	
	From the inverse Fourier transformation of Eq.~\eqref{eq:pressure-fourier}, we derive the expression for the pressure field in real space. It can be expressed as
	\begin{equation}
		p = \frac{V \nu_3}{a^2} \left( p^\mathrm{IB} + p^* \right) \, , 
	\end{equation}
	with $p^\mathrm{IB}$ denoting the solution for an isotropic Brinkman fluid, given by~\cite{martin2019two}
	\begin{equation}
		p^\mathrm{IB} = \frac{{c}^3 q_3}{r q_1} \, \cos\theta \, .
	\end{equation}
	
	The leading-order correction to the pressure field  only involves terms of the form $\cos(n\theta)$, with $n \in \{ 1,3,5,7\}$. Again, corresponding expressions are not explicitly included here due to their lengthiness. 
	However, the Maple script, which furnishes the final expressions for the pressure field, is available through the Zenodo repository.


	\section{Resistance coefficients}
	\label{sec:resistance}

	Having derived the leading-order corrections to the hydrodynamic fields for a disk translating in a homogeneous, weakly anisotropic fluid with friction, our next step is to quantify the correction to leading order in both $A$ and $\beta$ to the resistance coefficient of the disk in translational motion through the fluid.
	The frictional force exerted on the translating disk can be obtained by integrating the viscous traction at the surface of the disk.
	Specifically, the component of this so-called resistance force along the director is calculated as
	\begin{equation}
		F_\parallel = -\int_0^{2\pi} \left( \sigma_{\rho\rho} \cos\theta - \sigma_{\rho\theta} \sin\theta\right) \, a \, \mathrm{d}\theta
	\end{equation}
	for motion parallel to the director. 
	Its component perpendicular to the director is obtained via
	\begin{equation}
		F_\perp = -\int_0^{2\pi} \left( \sigma_{\rho\rho} \sin\theta + \sigma_{\rho\theta} \cos\theta \right) \, a \, \mathrm{d}\theta \, .
	\end{equation}
	In the linear world, these two contributions simply superimpose, depending on the actual direction of motion.
	Hence, we can determine the resistance force exerted on a disk translating in any arbitrary direction.
	
	Defining 
	\begin{align}
		\Pi_{\rho\rho} &= -\frac{p}{\nu_3} + 2  v_{\rho, \rho} \, , \\
		\Pi_{\rho\theta} &= v_{\theta, \rho} + \frac{v_{\rho, \theta} - v_\theta}{\rho} \, , 
	\end{align}
	{the frictional forces can be expressed as}
	\begin{subequations}
		\begin{align}
			F_\parallel &= a \nu_3 \int_0^{2\pi} \left. \left( \Pi_{\rho\theta} \cos\theta
			- \Pi_{\rho\rho} \sin\theta + {\pi_\parallel} \right) \right|_{\rho = a} \mathrm{d} \theta \, , \\
			F_\perp &= a \nu_3 \int_0^{2\pi} \left. \left( \Pi_{\rho\rho} \sin\theta + \Pi_{\rho\theta} \cos\theta + {\pi_\perp} \right) \right|_{\rho = a} \mathrm{d} \theta \, .
		\end{align}
	\end{subequations}
	In these expressions, we abbreviate
	\begin{widetext}
		\begin{align}
			{\pi_\parallel} &= \epsilon_+ \left(    v_{\rho, \rho} \left( 3 \cos\theta + \cos (3\theta) \right)
			-\Pi_{\rho\theta} \left( \sin\theta + \sin (3\theta) \right)
			+  
			\frac{v_\rho + v_{\theta, \theta}}{\rho} \left( \cos\theta - \cos (3\theta) \right) \right) , \\[3pt]
			{\pi_\perp} &= \epsilon_-
			\left( 
			v_{\rho, \rho} \left( 3 \sin\theta - \sin (3\theta) \right)
			+\Pi_{\rho\theta} \left( \cos\theta - \cos (3\theta) \right)
			+  
			\frac{v_\rho + v_{\theta, \theta}}{\rho} \left( \sin\theta + \sin (3\theta) \right)
			\right) ,
		\end{align}
	\end{widetext}
	where
	\begin{equation}
		\epsilon_\pm = \frac{A \pm 2b}{8} \, .
	\end{equation}
	
	We express the general solution for the hydrodynamic fields in the form of Fourier series as
	\begin{subequations}
		\begin{align}
			v_\rho^\parallel &= {V_\parallel} \sum_{n = 0}^\infty g_n^\parallel (\rho) \cos \big( (2n+1)\theta \big) \, , \\
			v_\theta^\parallel &= {V_\parallel} \sum_{n = 0}^\infty h_n^\parallel (\rho) \sin \big( (2n+1)\theta \big) \, , \\
			p^\parallel &= {\frac{V_\parallel \nu_3}{a^2}} \, \sum_{n = 0}^\infty w_n^\parallel (\rho) \cos \big( (2n+1)\theta \big) \, 
		\end{align}
	\end{subequations}
	for motion parallel to the director.
	Similarly, for motion perpendicular to the director, the sine and cosine are interchanged in these expressions, along with the utilization of the radial functions $g_n^\perp$, $h_n^\perp$, and $w_n^\perp$.
	{In particular, for $\beta = 0$, we find that \( g_n^\perp = (-1)^n g_n^\parallel \) and \( h_n^\perp = (-1)^{n+1} h_n^\parallel \); however, \( w_n^\perp = (-1)^n w_n^\parallel \) only when also $b=0$.}
	It follows from the boundary conditions imposed at the surface of the disk that $g_0^\parallel = g_0^\perp = h_0^\perp = 1$ and that $h_0^\parallel = -1$.
	In addition, it follows from the incompressibility equation that $\mathrm{d}g_i^\parallel/ \mathrm{d}\rho = \mathrm{d}g_i^\perp/ \mathrm{d}\rho = 0$ for $i \in \{0,1 \}$ at $\rho = a$.
	Accordingly, the resistance forces are obtained as
	\begin{subequations}
		\begin{align}
			\frac{F_\parallel}{\pi \nu_3 V_\parallel} &=
			\left. w_0^\parallel + a \, \frac{\mathrm{d}}{\mathrm{d}\rho} \left( \left( 1+\epsilon_+\right)h_0^\parallel + \epsilon_+ h_1^\parallel  \right) \right|_{\rho = a} , \\
			\frac{F_\perp}{\pi \nu_3 V_\perp} &=
			\left. w_0^\perp - a\, \frac{\mathrm{d}}{\mathrm{d}\rho} \left( \left( 1+\epsilon_-\right)h_0^\perp - \epsilon_- h_1^\perp  \right) \right|_{\rho = a} .
		\end{align}
	\end{subequations}

	We now introduce the hydrodynamic resistance coefficients relating the translational velocities to the corresponding components of the resistance force via $F_\parallel = -R_\parallel V_\parallel$ and $F_\perp = -R_\perp V_\perp$.
	After some algebraic manipulation, which involves expressing $w_0^\parallel$ and $w_0^\perp$ in terms of velocity as dictated by the governing equations of fluid motion, {we obtain
		\begin{equation}
			\frac{-R_\parallel}{\pi \nu_3}
			= \left.  \left(a\alpha_\parallel \right)^2
			+ a\, \frac{A}{4} \frac{\mathrm{d} h_1^\parallel}{\mathrm{d}\rho}
			+ a^2 \left( 1+\frac{A}{8} \right) \frac{\mathrm{d}^2 h_0^\parallel}{\mathrm{d} \rho^2}
			\right|_{\rho=a}  \label{eq:FORCE_PARA}
		\end{equation}
		for motion parallel to the director and
		\begin{equation}
			\frac{-R_\perp}{\pi \nu_3}
			= \left. \left(a \alpha_\perp \right)^2
			+ a \, \frac{A}{4} \frac{\mathrm{d} h_1^\perp}{\mathrm{d}\rho}
			- a^2 \left( 1+\frac{A}{8} \right) \frac{\mathrm{d}^2 h_0^\perp}{\mathrm{d} \rho^2}
			\right|_{\rho=a}  \label{eq:FORCE_PERP}
		\end{equation}
	}
	for motion perpendicular to the director.
	We mention that Eqs.~\eqref{eq:FORCE_PARA} and \eqref{eq:FORCE_PERP} {in their given form} do not impose any further restriction on the magnitude of the values of $A$ and~$\beta$.
	{We assume small magnitudes of  \( A \) and \( \beta \) to derive approximate expressions for \( h_0^\perp \), \( h_0^\parallel \), \( h_1^\perp \), and \( h_1^\parallel \). 
		These expressions, in turn, provide leading-order approximations for the resistance coefficients.}

	Particularly, for $\beta = 0$, we have $h_0^\perp = -h_0^\parallel$ and $h_1^\perp = h_1^\parallel$. Thus, we infer that, in this particular scenario, the resistance coefficient remains independent of the orientation of the director to the considered leading order in $A$ and $\beta$.
	Alternatively, the total resistance force exerted on the translating disk can be expressed as
	\begin{equation}
		F_k = -\langle F \rangle + \pi \left( a \alpha_k \right)^2 \nu_3 V_k \, 
	\end{equation}
	for $k \in \{\parallel, \perp\}$, with $\langle F \rangle = 2\pi a f_0$ denoting the force density averaged over the perimeter of the disk.

	Generally, the resistance coefficient for motion along the director to leading order in $A$ and $\beta$ can be calculated and expressed in the form
	\begin{equation}
		R_\parallel = \nu_3 \left( R^\mathrm{IB} + R_\parallel^A A + R_\parallel^\beta \beta \right) \, , 
	\end{equation}
	with $R^\mathrm{IB}$ representing the resistance coefficient of a disk moving in an isotropic Brinkman fluid~\cite{ramachandran2010drag,ota2018three}
	\begin{equation}
		R^\mathrm{IB} = \pi {c} 
		\left( {c} + 4\, \frac{q_2}{q_1} \right) \, .
		\label{eq:R_IB}
	\end{equation}
	In addition,
	\begin{equation}
		R_\parallel^A = \frac{\pi {c} }{4}
		\left( {c} + 2\, \frac{q_2}{q_1} - {c} \, \frac{q_2^2}{q_1^2} \right), 
		\label{eq:R_para_A}
	\end{equation}
	and
	\begin{equation}
		R_\parallel^\beta = -\pi \left( {c} \, \frac{q_2}{q_1} \right)^2 \, .
		\label{eq:R_para_beta}
	\end{equation}

	The hydrodynamic mobility, defined as the inverse of the resistance coefficient, is likewise obtained to leading order in $A$ and~$\beta$. It can be cast into the form
	\begin{equation}
		\mu_\parallel = \frac{\mu^\mathrm{IB}}{\nu_3} \left( 1 - k_\parallel^A A - k_\parallel^\beta \beta \right) \, , 
		\label{eq:mobi}
	\end{equation}
	wherein $\mu^\mathrm{IB} = 1/R^\mathrm{IB}$ is the hydrodynamic mobility in an isotropic Brinkman fluid medium.  $k_\parallel^A = R_\parallel^A/R^\mathrm{IB}$ and~$k_\parallel^\beta = R_\parallel^\beta/R^\mathrm{IB}$ are the leading-order corrections resulting from anisotropy. 
	In the limit of ${c} =\alpha_\parallel a \to 0$, {hence, \( \alpha_\perp = (1-\beta) \alpha_\parallel \to 0 \) due to Eq.~\eqref{eq:beta_def},} we obtain to leading order
	\begin{equation}
		\mu_\parallel = \frac{1}{4\pi\nu_3} 
		\left( \Lambda - \frac{2\Lambda-1}{16} \, A + \frac{\beta}{4} \right) 
		+ \mathcal{O} \left( {c}^2 \right) ,
		\label{eq:mu_parallel_c0}
	\end{equation}
	where we have defined
	\begin{equation}
		\Lambda = \ln \left( \frac{2}{{c}} \right) -\gamma \, ,
	\end{equation}
	with $\gamma$ denoting Euler’s constant.
	{Physically, this limit corresponds to a situation of weak friction with the substrate or the surrounding environment.}
	Conversely, in the opposite limit of ${c}=\alpha_\parallel a \to \infty$, 
	{hence, \( \alpha_\perp = (1-\beta) \alpha_\parallel \to \infty \),}
	{corresponding to a scenario of significant friction with the substrate or the surrounding environment,} we find
	\begin{equation}
		\mu_\parallel = \frac{1+\beta}{\pi \nu_3 } \, c^{-2} +  \mathcal{O} \left( c^{-3} \right) . \label{eq:mobi_c_inf}
	\end{equation}

	Next, we examine the motion of the disk in the direction perpendicular to the director. We find that $R_\perp^A = R_\parallel^A$.
	In addition,
	\begin{equation}
		R_\perp^\beta = \pi c^2 \left( 2 - 3 \, \frac{q_2^2}{q_1^2} \right) .
		\label{eq:R_perp_beta}
	\end{equation}
	
	In the asymptotic limit of $c =  \alpha_\parallel a \to 0$, we obtain
	\begin{equation}
		\mu_\perp = \frac{1}{4\pi\nu_3} 
		\left( \Lambda - \frac{2\Lambda-1}{16} \, A + \frac{3}{4}\, \beta \right) 
		+ \mathcal{O} \left( {c}^2 \right) ,
		\label{eq:mu_perp_c0}
	\end{equation}
	from where the coefficients $k^A_\perp$ of $A$ and $k^\beta_\perp$, defined in analogy to Eq.~(\ref{eq:mobi}) can be read off.
	The limit of $c=  \alpha_\parallel a\to \infty$ leads to the same result for $\mu_\perp$ as on the right-hand side of Eq.~\eqref{eq:mobi_c_inf}.

	\begin{figure}
		\centering
		\includegraphics[width=\columnwidth]{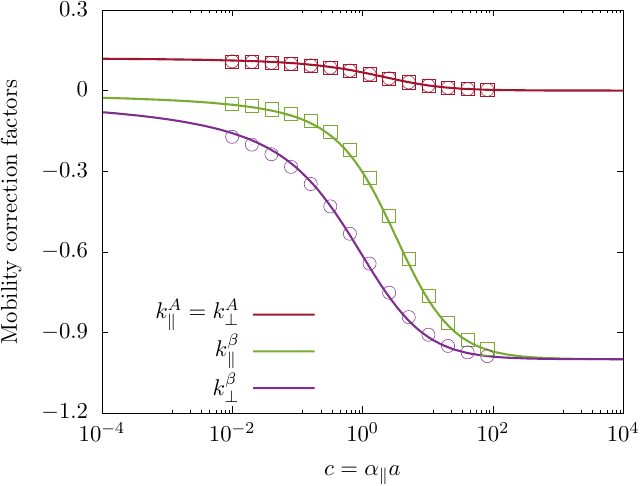}
		\caption{
			Variations of the correction factors of the hydrodynamic mobility function, denoted by $k_i^j = R_i^j / R^\mathrm{IB}$, where $i \in \{ \parallel, \perp \}$ and $j \in \{ A, \beta \}$. Solid lines depict analytical results derived from a perturbative expansion to leading order in $A$ and $\beta$. $R_\parallel^A = R_\perp^A$, $R_\parallel^\beta$, and $R_\perp^\beta$ are given by Eqs.~\eqref{eq:R_para_A}, \eqref{eq:R_para_beta}, and \eqref{eq:R_perp_beta} respectively. Symbols correspond to results inferred from finite-element simulations.
			The exact drag coefficient for an isotropic Brinkman fluid $R^\mathrm{IB}$ is given by Eq.~\eqref{eq:R_IB}.
		}
		\label{fig:mobi}
	\end{figure}

	\begin{figure}
		\centering
		\includegraphics[scale=0.45]{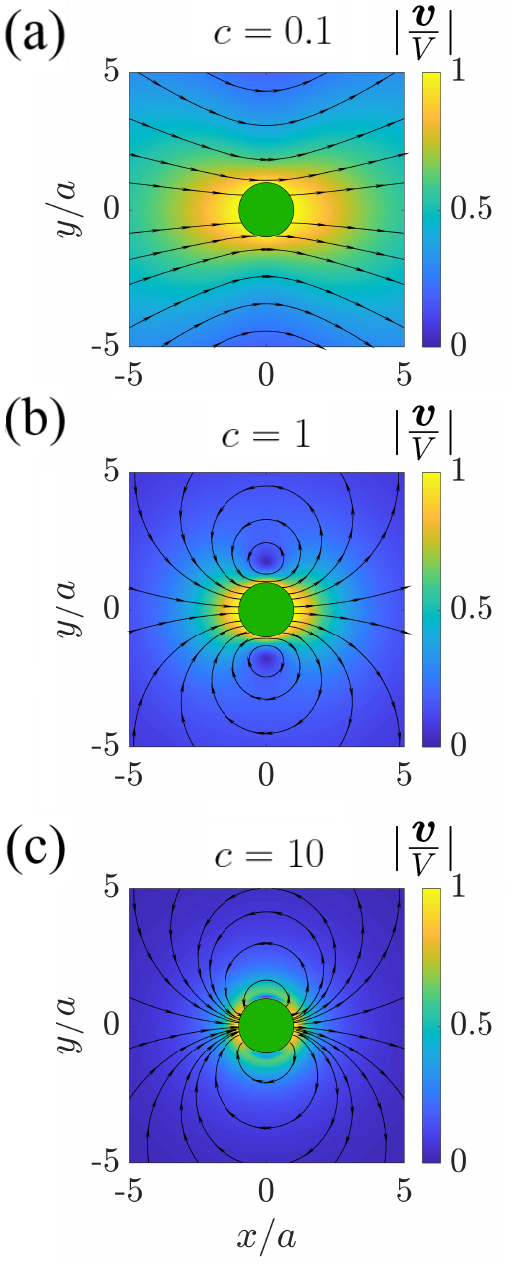}
		\caption{Streamlines and quiver plots depicting the resulting fluid flows when a disk is dragged through a two-dimensional, isotropic Brinkman fluid towards the positive $x$-direction. The different plots are showcasing situations of (a) $c=  \alpha_\parallel a= 0.1$, (b) $c=  \alpha_\parallel a= 1$, and (c) $c=  \alpha_\parallel a=10$. Analytical results are presented in the top half of each panel ($y>0$), while FEM results are provided in the bottom half ($y<0$). }
		\label{fig:IB}
	\end{figure}

	\begin{figure*}
		\centering
		\includegraphics[scale=0.5]{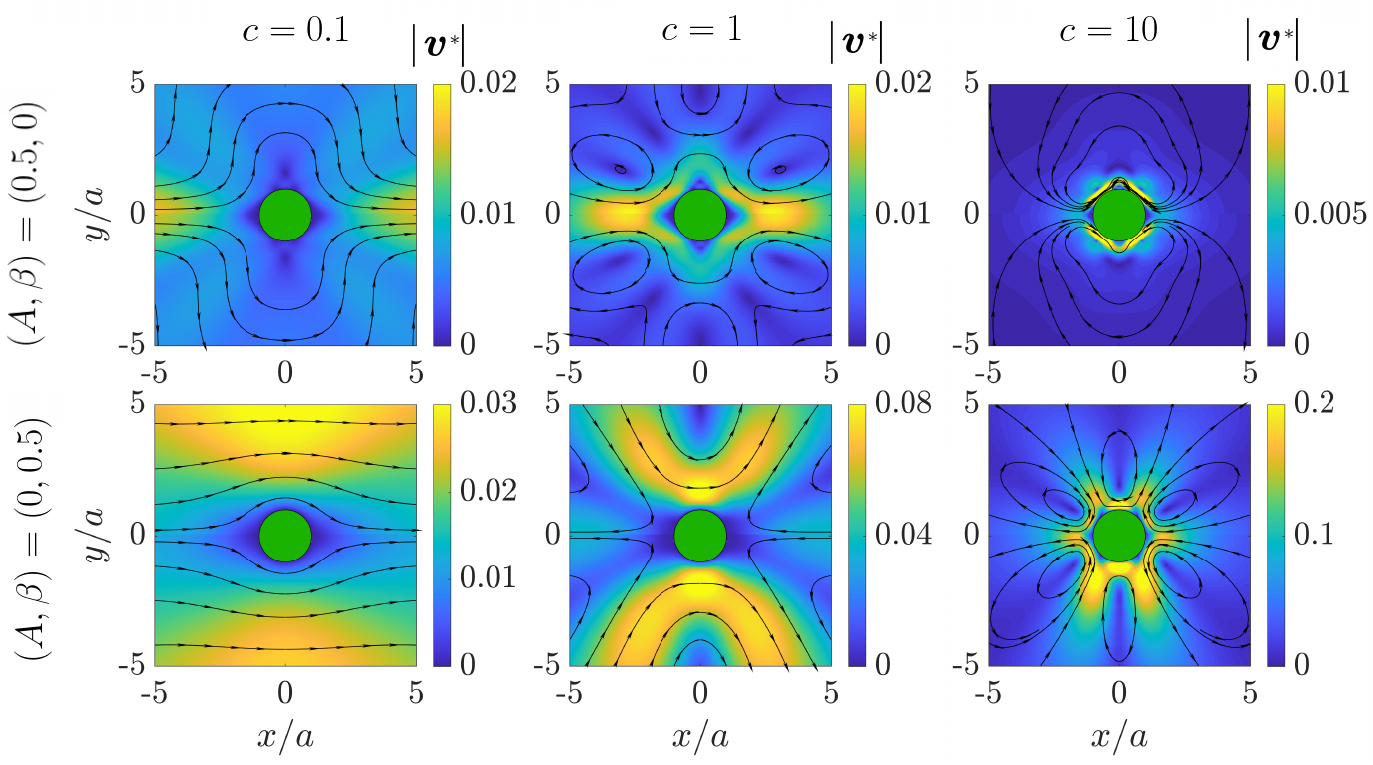}
		\caption{Streamlines and quiver plots illustrating the image flow field $\vect{v}^*$ induced by a disk translating at speed $V$ in a two-dimensional, uniaxially anisotropic Brinkman fluid. The disk moves along the axis of anisotropy into the positive $x$-direction. 
			Results are presented for three values of the parameter $c=\alpha_\parallel a$, namely $c=0.1$ (left column), $c=1$ (middle column), and $c=10$ (right column). Additionally, we consider values of the anisotropy parameters $A$ and $\beta$ as $(A,\beta) = (0.5, 0)$ (top row) and $(A,\beta) = (0, 0.5)$ (bottom row). Analytical solutions are depicted in the top halves ($y>0$), while FEM results are shown in the bottom halves ($y<0$). }
		\label{fig:para}
	\end{figure*}

	\begin{figure*}
		\centering
		\includegraphics[scale=0.49]{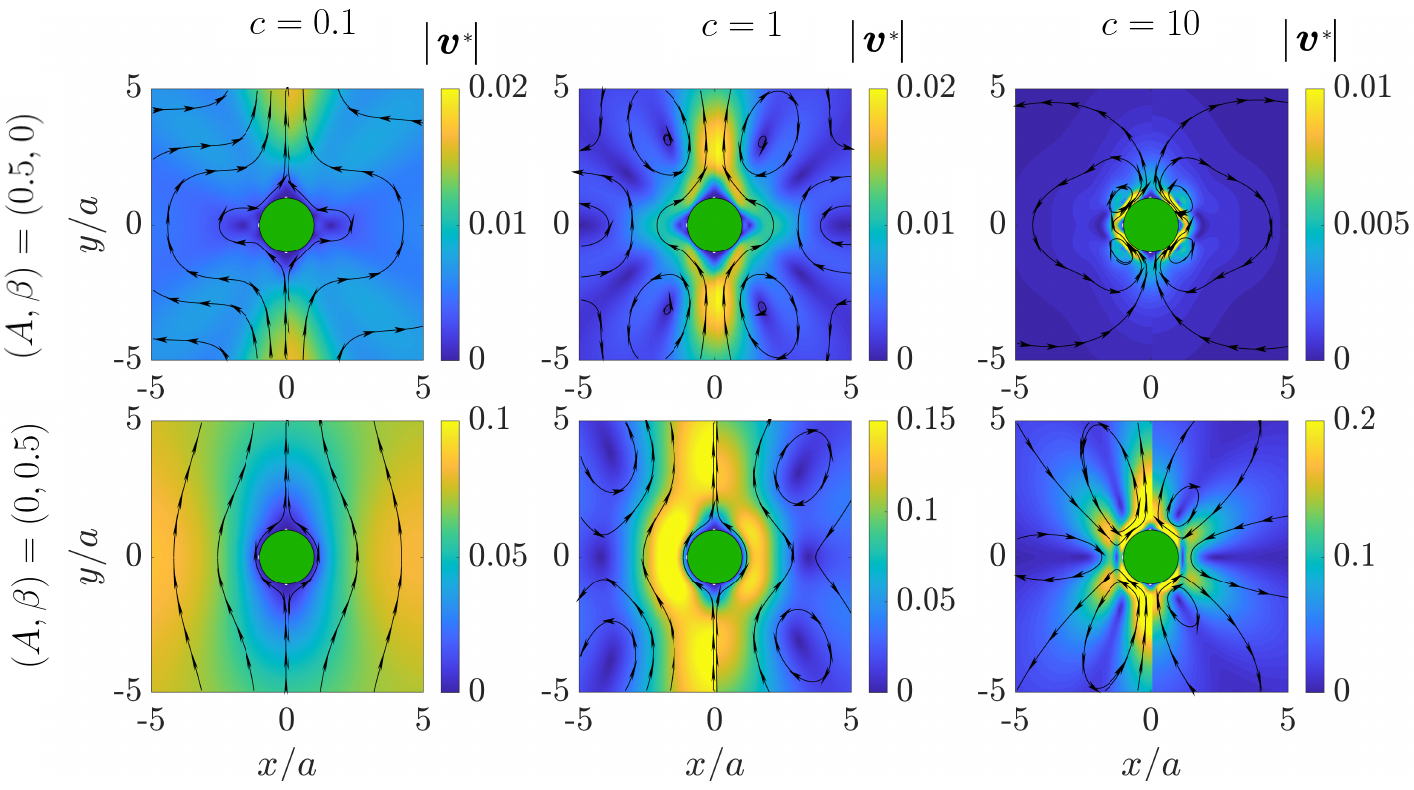}
		\caption{Streamlines and quiver plots illustrating the image flow field $\vect{v}^*$ for a disk translating perpendicular to the director in a two-dimensional, uniaxially anisotropic Brinkman fluid, here into the positive $y$-direction. Results are displayed for three values of the parameter $c=\alpha_\parallel a$, namely for $c=0.1$ (left column), $c=1$ (middle column), and $c=10$ (right column). Moreover, two different combinations of the anisotropy parameters $A$ and $\beta$ are considered, namely $(A,\beta) = (0.5, 0)$ (top row) and $(A,\beta) = (0, 0.5)$ (bottom row). We depict analytical solutions in the right-hand halves ($x>0$) and FEM results in the left-hand halves ($x<0$).}
		\label{fig:perp}
	\end{figure*}

	\begin{figure}
		\centering
		\vspace{-0.2cm}
		\includegraphics[scale=0.45]{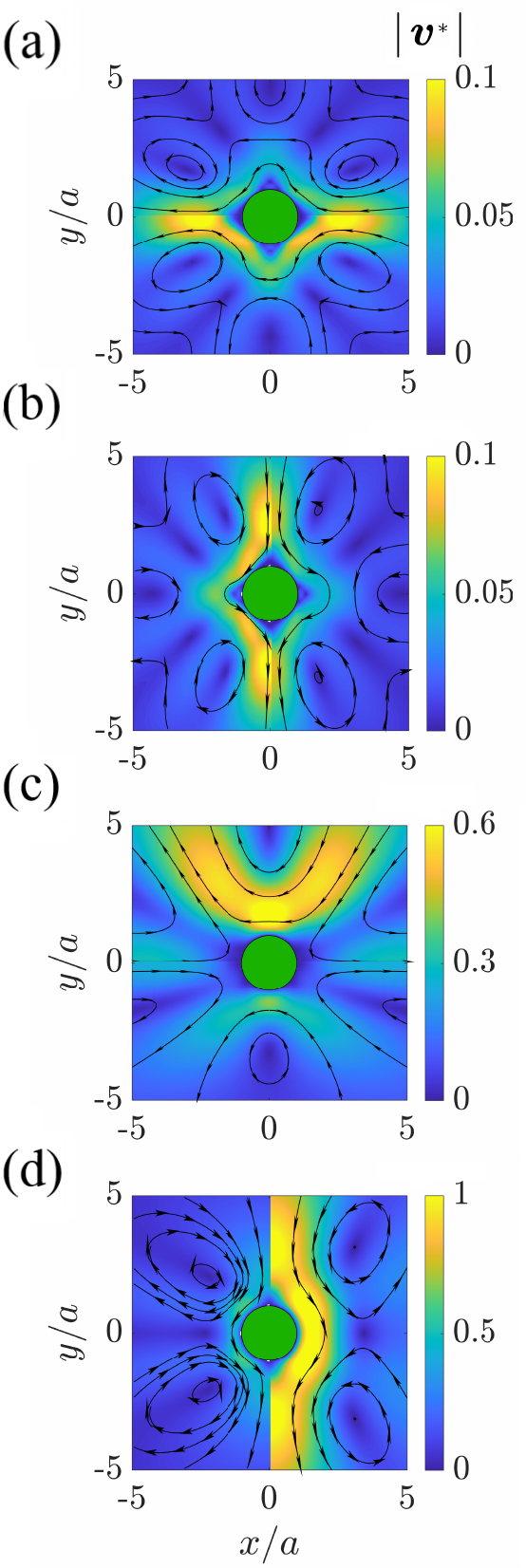}
		\vspace{-0.2cm}
		\caption{Streamlines and quiver plots illustrate the image flow field $\vect{v}^*$ induced by a translating disk in a two-dimensional, uniaxially anisotropic Brinkman fluid. The motion here is parallel to the director in panels (a) and (c), which include theoretical solutions in the top halves ($y>0$) and FEM results in the bottom halves ($y<0$). In panels (b) and (d) motion is perpendicular to the director, which includes theoretical solutions on the right-hand sides ($x>0$) and FEM results on the left-hand sides ($x<0$). We set $c=\alpha_\parallel a=1$. Here, the anisotropy parameters $A$ and $\beta$ are set to $(A,\beta) = (-2,0)$ in (a) and (b), while $(A,\beta) = (0, -4)$ in (c) and (d). }
		\label{fig:para-perp-extra}
	\end{figure}

	In Fig.~\ref{fig:mobi}, we illustrate the variations of the correction factors with respect to $c=  \alpha_\parallel a$, as derived above theoretically (solid lines) and as determined from finite-element simulations (symbols), which we outline below. Squares and circles represent the results obtained for motion parallel and perpendicular to the director, respectively.
	The simulations validate the equality $k_\parallel^A = k_\perp^A$, demonstrating the independence of the drag coefficient on the direction of motion. 
	Furthermore, it is evident that the correction attributed to anisotropy in friction, quantified by $\beta$, outweighs that of anisotropy in viscosity, quantified by $A$.
	{This behavior is far from trivial. 
		One conceivable reason lies with the nature  of two-dimensional fluid flows themselves. Generally, in an isotropic fluid, the Green's function quantifying the flows induced by a locally applied net force diverges logarithmically with the distance from the point of force application \cite{richter22,lutz22}. It is the linear friction with the environment that regularizes this divergence in Brinkman fluids so that the induced flows decay towards zero with infinite distance \cite{ramachandran2010drag}. This subtle role stresses the importance of the effect of friction in general for two-dimensional fluid flows.}
	Additionally, we observe that the impact of the anisotropy onto friction when compared to the isotropic case to leading order is more pronounced for motion perpendicular than parallel to the director, {at least in an intermediate regime}.
	{As inferred from Eq.~\eqref{eq:mobi}, the sign of the hydrodynamic mobility is determined by the expression $1 - k_i^A A - k_i^\beta \beta$, where $i \in \{\parallel, \perp\}$. Given that $A$ and $\beta$ are assumed to be much less than 1, this quantity is always positive.}

	\section{Finite-element simulations}
	\label{sec:FEM}

	To validate the analytical results, we carry out numerical simulations using the Gascoigne 3d finite element library~\cite{Braack2021}. 
	For this purpose, the set of Eqs.~(\ref{eq:Stokes}) and (\ref{eq:incompressibilityEqn}) 
	is formulated for $\vect{f}(\vect{r})=\vect{0}$ in variational formulation. Both fields, velocity $\vect{v}(\vect{r})$ and pressure $p(\vect{r})$ are discretized using quadratic finite elements. As this equal-order finite element pair is not satisfying the inf-sup condition, the discretization is stabilized with local projections~\cite{BeckerBraack2001}
	\[
	\vect{\nabla}\cdot\vect{v}  - \sum_{T\in\Omega_h} \delta_T \Delta (p-\pi^{(1)}p)=0,
	\]
	where $\pi^{(1)}$ is the projection to the space of linear finite elements with the parameter $\delta_T =  ( {\alpha_\parallel} +h_T^{-2})^{-1}$ that depends on  the local grid size $h_T$ and on ${\alpha_\parallel}$.
	
	To approximate the open boundary problem we introduce the numerical domain
	\begin{eqnarray}
		\Omega_L &=& \{-L<x<L,\; -L<y<L\}\setminus B_1,
		\nonumber\\ 
		B_1 &=& \{x\in\mathbb{R}^2,\; \|x\|_2<1\}.    
	\end{eqnarray}
	We choose $L$ large enough so that the solution decays sufficiently in magnitude when reaching the boundaries. We thus assume $\vect{v}=\vect{0}$ as Dirichlet conditions on the outer boundary. On the disk, $\bm{v}=(V,0)$  or $\bm{v}=(0,V)$ is enforced. For $L=200$, the artificial truncation of the domain is no longer perceptible as long as ${\alpha_\parallel} \ge 0.1$. For smaller values of ${\alpha_\parallel}$, we must choose $L$ even larger. To keep computing time low, we use local mesh refinement and only refine elements within the square $[-10,10]^2$. The local mesh size at the disk is about $h_T\approx 0.015$. The resulting linear problems are solved with a GMRES solver~\cite{saad86}, preconditioned by a parallel multigrid method~\cite{FailerRichter2019}. Computing times are just a few seconds per configuration.
	
	To evaluate the drag on the disk, we integrate the viscous stress along the perimeter of the disk, taking into account the direction of motion. 
	We use the Babu\v{s}ka-Miller trick~\cite{BabuskaMiller1984a} to express these line integrals via volume integrals. 
	%
	%
	This formulation gives the drag with highest accuracy and fourth-order convergence in the mesh size $h_T$, see Refs.~\onlinecite{J2} and \onlinecite[(Remark 8.17)]{Richter2017} for further details.

	\section{Results and discussion}
	\label{sec:res_discussion}
	
	At this point, we add a few additional remarks. They mainly concern the two-dimensional character of the fluid that becomes apparent from our results. Specifically, it concerns the expressions of $\mu_\parallel$ and $\mu_\perp$ for $\alpha_\parallel\rightarrow0$ and $\alpha_\perp\rightarrow0$, respectively. That is, linear friction tends to zero. Thus, the fluid flow in Eq.~(\ref{eq:Stokes}) is hardly dampened from outside. As apparent from Eqs.~(\ref{eq:mu_parallel_c0}) and (\ref{eq:mu_perp_c0}), respectively, the two mobilities in this case diverge logarithmically for $\alpha_\parallel\rightarrow0$ and $\alpha_\perp\rightarrow0$. 
	
	This behavior is in line with the features observed already for an isotropic fluid in a two-dimensional setting without external damping. In this situation, the magnitude of the resulting velocity field diverges logarithmically with the distance from the position where a net force is imposed onto the fluid \cite{richter22,lutz22}. It manifests that the infinitely extended two-dimensional sheet of fluid is not sufficient to stabilize the flow in that case. In contrast to the situation of a three-dimensional bulk system, in which the infinite amount of fluid stabilizes the flow, in the two-dimensional case the whole fluid is set into motion under persistent forcing and a stationary situation does not exist. The flow in two dimensions can be stabilized by adding appropriate counterforces on the fluid so that the whole forcing sums up to zero \cite{richter22} and/or by introducing regularizing boundaries \cite{lutz22}.

	In Fig.~\ref{fig:IB}, we present streamlines and quiver plots illustrating the behavior of a disk translating in a two-dimensional, isotropic Brinkman fluid. It moves at the origin into the positive $x$-direction. Results are depicted in the laboratory frame of reference for three different values of $c=\alpha_\parallel a$, namely (a) $c = 0.1$, (b) $c = 1$, and (c) $c=10$. Due to the symmetry of the set-up, the induced flow fields are symmetric with respect to the $x$-axis, that is, with respect to $y=0$. Therefore, we display for comparison analytical results in the upper half of the plots ($y>0$) in Fig.~\ref{fig:IB}, while FEM results are depicted in the lower half ($y<0$).
	Naturally, as the friction coefficients are increased, there is a notable decline in the magnitude of the flow field. This is in line with an effective screening of the flow induced by the disk due to the frictional interaction with the surroundings. At the intermediate value of $c$ in Fig.~\ref{fig:IB}(b), circulating flows become apparent in the vicinity of the translating disk. 
	Besides illustration, the comparison in Fig.~\ref{fig:IB} confirms the simulations performed using the finite-element method (FEM). We subsequently utilize it to test the theoretical predictions for a weakly anisotropic fluid medium by FEM simulations.
	
	We thus proceed to assess the adequacy of the perturbative expansion in the analytical theory through rigorous comparison with FEM simulations. To focus on the deviation from isotropic Brinkman effects in scenarios of weak anisotropy, we showcase the image flow field $\vect{v}^*$ rather than the complete flow field $\vect{v}$.
	In Figs.~\ref{fig:para}, we present streamlines and quiver plots illustrating $\vect{v}^*$, for motion parallel to the director, that is, into the $x$-direction. 
	Results are presented for three cases, namely for $c=0.1$ (left column), $c=1$ (middle column), and $c=10$ (right column). Additionally, the parameters $A$ and $\beta$ are varied as $(A,\beta) = (0.5, 0)$ (top row) and $(A,\beta) = (0, 0.5)$ (bottom row). 
	Again, we present the analytical solutions in the top halves of the plots ($y>0$), while FEM results are shown for comparison in the bottom halves ($y<0$).
	
	Overall, the remarkable agreement between theory and FEM simulations underscores the effectiveness of the employed theoretical approach, at least for the considered moderate values of the anisotropy parameters $A$ and $\beta$. Notably, the image solution occasionally reveals regions of vortex flows, characterized by closed streamlines.
	
	Next, in Fig.~\ref{fig:perp}, we turn ourselves to motion of the disk perpendicular to the axis of anisotropy. That is, we assume motion into the positive $y$-direction. The same parameter values as considered in Fig.~\ref{fig:para} are investigated.
	Due to the perpendicular direction of motion, we here present analytical solutions in the halves of the plots on the right-hand sides ($x>0$), while FEM results are displayed in the halves on the left-hand sides ($x<0$).
	We observe a notable agreement between the two halves of each plot. The agreement is particularly evident for scenarios of low magnitudes of $c=\alpha_\parallel a$ (left column) or for $\beta = 0$ (top row).
	However, for higher values of $c$ combined with $\beta \neq 0$, 
	the analytical theory appears to underestimate the magnitudes of the velocity field, while qualitatively there is still good agreement. In other words, the overall structure of the flow fields and the configurations of the streamlines exhibit considerable consistency across all cases examined.

	In Fig.~\ref{fig:para-perp-extra} we challenge the theory by examining values of $A$ and $\beta$ that we may assume are beyond the range of validity of the perturbative expansion. We keep $c=\alpha_\parallel a$ fixed at the magnitude $c=1$. 
	For $(A,\beta) = (-2,0)$ as depicted in panels (a) and (b), the theory tends to underestimate the magnitude of the velocity field $\vect{v}^*$ when compared to FEM results. We remark that for negative values of $A$ and $\beta$, the perturbative solution for the image flow field $\vect{v}^*$ at the leading order provides the same flow field, except for an opposite sign. Both, translation of the disk (a) parallel and (b) perpendicular to the axis of uniaxiality are considered, with theoretical solutions at the top and on the right-hand side, while FEM results are included at the bottom and on the left-hand side, respectively. 
	The same applies to panels (c) and (d), respectively, where we illustrate the case of $(A,\beta) = (0, -4)$. 
	There, the theory overestimates the magnitude of the velocity field $\vect{v}^*$ when compared to FEM results.
	Still, in all considered cases, there is convincing qualitative agreement concerning the structure of the flow fields and streamlines. 
	Enhanced quantitative agreement between theory and simulations could potentially be achieved by including higher-order terms in the expansion concerning the anisotropy of the fluid, which may be explored in future research.
	
	{
		When comparing the resulting magnitudes of the flow field upon varying the anisotropy parameters $A$ and~$\beta$, we observe that the fluid velocity is significantly larger for nonzero values of $\beta$ than for nonzero values of $A$ of equal magnitude. Apparently, changes in $A$ have a less significant impact across the wide range of values of $c$ considered. This trend is also observed when we address the direction of motion. The magnitude of the velocity is larger for motion perpendicular to the director than parallel to it. These observations align with the previously discussed behavior regarding the contribution of each anisotropy parameter to the correction factors in the hydrodynamic mobilities, as long as $c=\alpha_\parallel a$ does not grow too large, see Fig.~\ref{fig:mobi}.}

\begin{figure}
		\centering
		\includegraphics[scale=0.51]{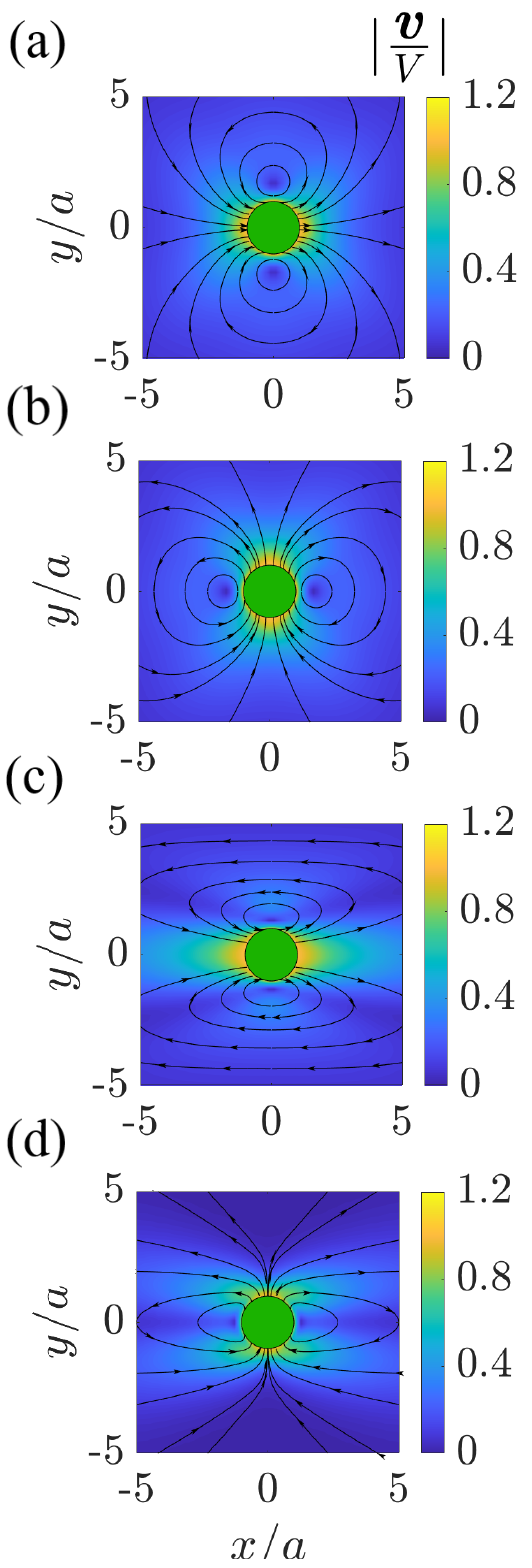}
		\caption{{Total flow field $\vect{v}$ for the same parameters as in Fig.~\ref{fig:para-perp-extra}, where the image flow field $\vect{v}^*$ was illustrated. The results on display are obtained from finite element simulations.}}
		\label{fig:para-perp-extra-full}
	\end{figure}

	{
		Finally, we present in Fig.~\ref{fig:para-perp-extra-full} the total flow field $\vect{v}$ (not only the image flow field $\vect{v}^*$) as obtained from finite element simulations using the parameters from Fig.~\ref{fig:para-perp-extra}. When compared to the flow in an isotropic medium in Fig.~\ref{fig:IB}~(b) for \(c=1\), we observe that for \((A, \beta) = (-2, 0)\), see Fig.~\ref{fig:para-perp-extra-full} (a) and (b), the structure of the flow field is not significantly affected by the anisotropy in viscosity. Conversely, the case \((A, \beta) = (0, -4)\) in Fig.~\ref{fig:para-perp-extra-full} (c) and (d) demonstrates a strong impact of the anisotropy in friction on the flow behavior and streamlines.
		Since in our investigations the contribution of the image solutions to the total flow field is generally small compared to the isotropic component, we do not observe the vortical structures when concentrating on the overall flow.
	}

	\section{Conclusions}
	\label{sec:conclusions}
	
	In summary, we investigate the translation of a circular disk in a two-dimensional layer of an incompressible, uniaxially anisotropic fluid that experiences linear friction under fluid flow. For instance, such friction could be due to the interaction with a supporting substrate. We confine ourselves to the regime of overdamped motion. The orientation of the anisotropy axis is assumed to be globally fixed. No-slip conditions apply for the fluid on the perimeter of the disk. 
	
	Using analytical theory, we calculate the resulting flows induced by the translating disk in the surrounding fluid. Moreover, we determine associated mobilities and resistance coefficients for the motion of the disk. To this end, we expand the theoretical description around the case of an isotropic fluid to leading order in the parameters of anisotropy. One of them quantifies the anisotropy of the viscosity of the fluid, the other one the anisotropy of the linear friction that the flow is exposed to. The flow fields determined by the theory are compared to results from finite-element simulations, with good quantitative agreement at moderate and still qualitative agreement at elevated magnitudes of the parameters of anisotropy. Interestingly, to leading order in the anisotropy of the viscosity, if the linear friction of the flow remains isotropic, the resistance coefficient are independent of the direction of motion. 
	
	The steady-state motion in two-dimensional fluids is generally an interesting topic due to logarithmic divergences arising in the theory, known as the Stokes paradox. In our situation, such divergences do not occur. The reason is found with the linear friction that additionally dampens the fluid flow. A related topic is given by displacements in two-dimensional elastic media \cite{richter22,lutz22}. 
	
	Overall, due to the specific features of two-dimensional flows, we hope to stimulate experimental confirmation. Introducing a globally ordered uniaxially anisotropic fluid layer may be achieved, for instance, in liquid-crystalline films on supporting substrates, if the substrate is rubbed accordingly to orient the nematic director. Incompressibility of the fluid within the plane can be realized by introducing an additional top coverslide to avoid varying thickness of the liquid-crystalline film. Another extension concerns the situation of thin elastic films and membranes. Our plan is to address membranes of uniaxial nematic elastomers \cite{urayama2007selected, menzel2007nonlinear, menzel2009response} as a next step.

	\appendix

	\section{List of parameters}

	\label{appendix:tab}
	
	{In Tab.~\ref{tab:parameters}, we summarize and provide an overview of the key parameters introduced in the main text for quick reference.}

	\begin{table}[]
		\centering
		{\renewcommand{\arraystretch}{1.5}
			{
				\begin{tabular}{|c|c|}
					\hline
					~~Parameter~~ & Definition \\
					\hline
					$\bar{\nu}$ & $2 \left( \nu_1 + \nu_2 - 2\nu_3 \right)$ \\
					$\zeta_1$ & $\left( 2\nu_1+\nu_2-\nu_4+\nu_5 \right)/\nu_3 - 1$ \\
					$\zeta_2$ & $\left( \nu_2+\nu_4-\nu_5\right)/\nu_3 - 1$ \\
					$\alpha_\parallel^2$ & $m_\parallel^2 / \nu_3$ \\
					$\alpha_\perp^2$ & $m_\perp^2 / \nu_3$ \\
					$A$ & $\bar{\nu}/\nu_3$ \\
					$B$ & $1 + A \sin^2\phi \cos^2\phi$ \\
					$H$ & $\alpha_\parallel^2 \sin^2\phi + \alpha_\perp^2 \cos^2\phi + k^2 B$ \\
					$w$ & $a/\rho$ \\
					$\sigma$ & ~~$\left. \rho
					\left( \alpha_\parallel^2 \sin^2\phi + \alpha_\perp^2 \cos^2\phi \right)^{1/2} \middle/ B^{1/2} \right. $~~ \\
					$\beta$ & $1 - \alpha_\perp / \alpha_\parallel$ \\
					$r$ & $\alpha_\parallel \rho$ \\
					$c$ & $\alpha_\parallel a$ \\
					$b$ & $ \left( \nu_1 - \nu_4 + \nu_5 \right)/\nu_3 $ \\
					$\epsilon_\pm$ & $\left( A \pm 2b \right)/8$ \\
					\hline
				\end{tabular}
			}
		}
		\caption{{List of the main parameters defined and used in the main text.} }
		\label{tab:parameters}
	\end{table}

	\section{Analytical expression of $\Phi_{mn}$}

	\label{appendix:int}
	
	{
		In the following, we provide the proof for the analytical result listed in Eq.~\eqref{eq:improper_int}. To begin, we define the integral
		\begin{equation}
			g(\sigma, w) =  \int_0^\infty \frac{x \, J_{2m}(x) J_{2n}({w} x)}{x^2 + \sigma^2} \,  \, \mathrm{d} x \, , 
			\label{eq:Phi_modified}
		\end{equation}
		with $m,n \in \mathbb{N}$, $\sigma \in \mathbb{R}$, and $w \in [0,1]$.
	}
	
	{
		Next, we introduce the Hankel function of the first kind of order $2m$ as~\cite{abramowitz72}
		\begin{equation}
			H_{2m}^{(1)} (z) = J_{2m}(z) + i Y_{2m}(z) \, .
		\end{equation}
		In this expression, \( Y_{2m} \) denotes the Bessel function of the second kind of order \( 2m \). The Hankel function of the first kind, \( H_{2m}^{(1)} \), has a branch cut along the negative real axis. On the upper side of this branch cut, for \( x > 0 \), we have
		\begin{equation}
			H_{2m}^{(1)} (-x) = -J_{2m}(x) + i Y_{2m}(x) \, .
		\end{equation}
		We recall that, for $x>0$,
		\begin{subequations}
			\begin{align}
				J_{\nu}(xe^{i \pi}) &= e^{i \pi \nu} J_{\nu}(x) \, , \\
				Y_{\nu}(xe^{i \pi}) &= e^{- i \pi \nu} Y_{\nu}(x) + 2i \cos(\pi \nu) J_{\nu}(x) \, .
			\end{align}
		\end{subequations}
		Then, since
		\begin{equation}
			\frac{x \, J_{2n}({w} x) }{x^2 + \sigma^2}  
		\end{equation}
		is an odd function of $x$, we find
		\begin{equation}
			g(\sigma, w) = \frac{1}{2} \operatorname{Re} \int_{-\infty}^\infty
			\frac{x \, H_{2m}^{(1)}(x) J_{2n}({w} x)}{x^2 + \sigma^2} \,  \, \mathrm{d} x \, ,
		\end{equation}
		where the integration from \(-\infty\) to 0 is taken along the upper side of the branch cut.
	}

	{
		Now, let us define the function $f(z)$ in the complex plane as 
		\begin{equation}
			f(z) = \frac{z \, H_{2m}^{(1)}(z) J_{2n}({w} z)}{z^2 + \sigma^2} \, .
		\end{equation}
		Performing a series expansion of \( f(z) \) around \( z = 0 \), we find that \(\lim_{z \to 0} f(z) = 0\) for \( m \le n \). Conversely, for \( m > n \), \( f(z) \) around \( z = 0 \) can be expressed as
		\begin{equation}
			f(z) = \sum_{k=-1}^{-m} a_{2k+1} z^{2k+1} + \mathcal{O}(z) \, . \label{eq:f_series}
		\end{equation}
		As there are no even negative power terms in the series expansion provided by Eq.~\eqref{eq:f_series}, we can express
		\begin{equation}
			g(\sigma, w) = \frac{1}{2} \, \operatorname{Re} \,
			\dashint_{-\infty}^\infty f(x) \, 
			\mathrm{d}x \, , 
		\end{equation}
		with the dashed integral representing the Cauchy principal value.
	}

	{
		Let us integrate the function \( f(z) \) along a contour composed of the upper side of the branch cut from \( -R \) to \( -\epsilon \), a small clockwise-oriented semicircle of radius \( \epsilon \) centered at the origin denoted by \( C_\epsilon \), the positive real axis from \( \epsilon \) to \( R \), and the counterclockwise path along the upper half of the circle \( |z| = R \).
	}
	
	{
		Leveraging that as \( z \to \infty \)
		\begin{subequations}
			\begin{align}
				J_\nu(z) &\sim \left( \frac{2}{\pi z} \right)^\frac{1}{2} 
				\cos \left( z - \frac{\nu \pi}{2} - \frac{\pi}{4} \right) \, , \\
				H_\nu^{(1)} &\sim \left( \frac{2}{\pi z} \right)^\frac{1}{2}
				e^{i \left( z - \frac{\nu \pi}{2} - \frac{\pi}{4} \right) } \, , 
			\end{align}
		\end{subequations}
		it follows that
		\begin{equation}
			|f(z)| \sim \frac{2}{\pi} \frac{1}{w^{\frac{1}{2}} |z|^2} \, \cos \left( z-\frac{\pi}{4} \right) ,
		\end{equation}
		as $|z| \to \infty$ in the upper half plane.
		Consequently, as \( R \) approaches infinity, the integral vanishes over the upper half of the circle \( |z| = R \).
	}

	{
		Then, as \( \epsilon \) tends to zero, the residue theorem yields
		\begin{equation}
			\dashint_{-\infty}^\infty f(x) \, 
			\mathrm{d}x + \lim_{\epsilon\to 0} 
			\int_{ C_\epsilon } f(z) \, \mathrm{d}z 
			= 2i\pi \operatorname{Res} \left\{ f(z), i\sigma \right\} .
			\label{eq:main_appendix}
		\end{equation}
	}

	{
		Using the connecting formulas~\cite{abramowitz72}
		\begin{subequations}
			\begin{align}
				I_\nu (z) &= e^{ \mp \frac{i\pi\nu}{2} } J_\nu
				\left( z e^{ \pm \frac{i\pi}{2} } \right) \, , \\
				K_\nu(z) &= \frac{i\pi}{2} \, e^{ \frac{i\nu\pi}{2} } H_\nu^{(1)} \left( z e^{ \frac{i\pi}{2} } \right) \, ,
			\end{align} \label{eq:connecting_formula}
		\end{subequations}
		the residue at \( z = i\sigma \) is calculated and expressed as
		\begin{equation}
			\operatorname{Res} \left\{ f(z), i\sigma \right\}
			= \frac{ (-1)^{m+n} }{2i\pi} \, 
			K_{2m}(\sigma) I_{2n} (w \sigma) \, .
			\label{eq:residue}
		\end{equation}
	}
	
	{
		Now, let us focus on the contour integral along the small clockwise-oriented semicircle of radius \( \epsilon \) centered at the origin. Since \( f(z) \) is an odd function, we can evaluate this integral using the residue theorem by considering a complementary semicircle of radius \( \epsilon \) in the lower plane.
		This approach leads to
		\begin{equation}
			\int_{ C_\epsilon } f(z) \, \mathrm{d}z
			= -i\pi \operatorname{Res} \left\{ f(z),0 \right\} \, , 
			\label{eq:int_inner_circle}
		\end{equation}
		The negative sign arises from the clockwise direction of the contour integration. Utilizing partial fraction decomposition, we obtain
		\begin{equation}
			\operatorname{Res} \left\{ f(z),0 \right\}
			=  
			\operatorname{Res} \left\{ h(z,-i\sigma), 0 \right\}
			+
			\operatorname{Res} \left\{ h(z,i\sigma), 0 \right\}\,,
		\end{equation}
		where
		\begin{equation}
			h(z, \alpha) = -\frac{1}{2} \frac{H_{2m}^{(1)}(z) J_{2n}(wz)}{\alpha-z} \, .
		\end{equation}
	}
	
	{
		From complex calculus, it is known that, if a function \(\phi(z)\) is analytic at \(z=\alpha\) and has a pole at \(z=a\), then the residue of \( \phi(z) / \left( \alpha-z\right) \) at \(z=a\) equals the principal part of the Laurent series expansion of \(\phi(z)\) about \(z=a\), evaluated at \(z = \alpha\).
		Accordingly, 
		\begin{align}
			\operatorname{Res} \left\{ f(z),0 \right\}
			&=  
			-\frac{1}{2} \, \mathcal{P} 
			H_{2m}^{(1)}(-i\sigma) J_{2n}(-iw\sigma) \notag \\
			&\quad-\frac{1}{2} \, \mathcal{P} 
			H_{2m}^{(1)}(i\sigma) J_{2n}(iw\sigma) \, .
		\end{align}
		Using the connecting formula~\cite{abramowitz72}
		\begin{equation}
			H_{\nu}^{(1)}(e^{- \frac{i\pi}{2} } z ) = 
			2e^{- \frac{i\nu\pi}{2} } I_{\nu}(z)- \frac{2i}{\pi} e^{\frac{i\nu\pi}{2} } K_{\nu}(z) \, 
		\end{equation}
		together with Eqs.~\eqref{eq:connecting_formula} and using $ \mathcal{P} I_{2m}(\sigma) I_{2n}(w\sigma) = 0 $, we obtain
		\begin{equation}
			\operatorname{Res} \left\{ f(z),0 \right\}
			= \frac{2}{\pi} \, (-1)^{m+n} \,
			\mathcal{P} K_{2m}(\sigma) I_{2n}(w\sigma) \, ,
			\label{eq:residue_at_zero}
		\end{equation}
		which is zero when $n \ge m$.
	}
	
	{
		Finally, we combine the results utilizing Eqs.~\eqref{eq:main_appendix}, \eqref{eq:residue}, \eqref{eq:int_inner_circle}, and \eqref{eq:residue_at_zero}. In this way, we arrive at the analytical solution presented in Eq.~\eqref{eq:improper_int}.
	}

	\section*{Data availability}
	
	The Maple~23 scripts containing the final expressions for the velocity and pressure fields are accessible through the Zenodo repository using the DOI \href{https://zenodo.org/records/10817265}{10.5281/zenodo.10817264}.
	
	\begin{acknowledgments}
		E.T.\ acknowledges the support received from the EPSRC under grant no.\ EP/W027194/1. A.M.M.\ acknowledges support provided by the Deutsche Forschungsgemeinschaft (DFG, German Research Foundation) through research grant no.\ ME 3571/5-1. Furthermore, A.M.M.\ thanks the DFG for support via the Heisenberg grant no.\ ME 3571/4-1.
	\end{acknowledgments}

	\bibliography{biblio}
\end{document}